\documentclass[prd,aps,reprint,nofootinbib,superscriptaddress,floatfix,onecolumn,eqsecnum]{revtex4-2}

\usepackage{todonotes}
\usepackage{marginnote}

\usepackage{listings}
\usepackage{graphicx}
\usepackage{dcolumn}
\usepackage{bm}
\usepackage{booktabs}
\usepackage{amsmath,amssymb}
\usepackage{subcaption}
\usepackage{framed}
\usepackage{xcolor}
\usepackage{physics}
\usepackage[colorlinks,linkcolor=blue]{hyperref}
\usepackage{orcidlink}
\usepackage{setspace}
\usepackage[normalem]{ulem}
\usepackage{empheq}
\usepackage{cancel}
\newcommand{\cO}{\mathcal{O}}
\newcommand{\cL}{\mathcal{L}}
\newcommand{\cA}{\mathcal{A}}

\newcommand{\tR}{\tilde{R}}

\newcommand{\tPhi}{{\tilde{\Phi}_0}}
\newcommand{\tphi}{\tilde{\varphi}_0}
\newcommand{\td}{\tilde{\delta}^\pm}
\newcommand{\tdp}{\tilde{\delta}^+}

\newcommand{\tom}{\tilde{\omega}^\pm}
\newcommand{\thg}{\theta_g}

\newcommand{\nn}{\nonumber\\}

\newcommand\tit{\textit}
\newcommand\mbf{\mathbf}
\newcommand\mcal{\mathcal}
\newcommand\mbb{\boldsymbol}

\newcommand\be{\begin{equation}}
\newcommand\bt{\begin{table}}
\newcommand\bts{\begin{table*}}
\newcommand\bfig{\begin{figure}}
\newcommand\bfs{\begin{figure*}}
\newcommand\bi{\begin{itemize}}
\newcommand\ee{\end{equation}}
\newcommand\et{\end{table}}
\newcommand\ets{\end{table*}}
\newcommand\efig{\end{figure}}
\newcommand\efs{\end{figure*}}
\newcommand\ei{\end{itemize}}
\newcommand{\ba}{\begin{align}}
\newcommand{\ea}{\end{align}}

\title{Your Paper}

\begin{document}

\title{Cosmological axion-photon interactions with a nonminimal dark gauge sector}
\author{Hao Jiao}
\email{jiaohao@ibs.re.kr}
\affiliation{Cosmology, Gravity and Astroparticle Physics Group, Center for Theoretical Physics of the Universe,
Institute for Basic Science (IBS), Daejeon, 34126, Korea}
\author{Dong-Won Jung}
\email{dongwon.jung@ibs.re.kr}
\affiliation{Cosmology, Gravity and Astroparticle Physics Group, Center for Theoretical Physics of the Universe,
Institute for Basic Science (IBS), Daejeon, 34126, Korea}
\author{Nils A. Nilsson}
\email{nilsson@ibs.re.kr}
\affiliation{Cosmology, Gravity and Astroparticle Physics Group, Center for Theoretical Physics of the Universe,
Institute for Basic Science (IBS), Daejeon, 34126, Korea}
\affiliation{LTE, Observatoire de Paris, Université PSL, CNRS, LNE, Sorbonne Universit\'e, 61 avenue de l’Observatoire, 75 014 Paris, France}

\date{\today}

\begin{abstract}
We study the cosmological evolution of radiation and dark radiation with nonminimal axion-photon-dark photon couplings. The Chern-Simons-type axion-gauge field interactions between the dark and visible sectors can induce energy conversion between photons and dark photons as well as polarization rotation during the propagation of radiation, leading to CMB spectral distortions and cosmic birefringence. Using the observed CMB monopole spectrum, we constrain the photon-dark photon energy conversion fraction, which is determined by the coupling constants and the evolution of the axion field. We further show that the parity-odd mixing interaction $\phi F\tilde{B}$ can partially relax the constraints on the standard axion-photon coupling, $\phi F\tilde{F}$, derived from the current observational limits of the cosmic birefringence angle.
\end{abstract}

\maketitle

\section{Introduction}
\label{sec:introduction}
The cosmic microwave background (CMB) constrains both the contents of the Universe and the evolution of its perturbations, and it allows us to test interactions which affect photons entering the horizon around the time of last scattering. Linear polarization is generated by Thomson scattering around recombination and during reionization~\cite{HuWhite1997}, and the resulting pattern can be decomposed into $E$ and $B$ modes of opposite parity~\cite{ZaldarriagaSeljak1997,KamionkowskiEtAl1997}. The ensemble-averaged $TB$ and $EB$ correlations vanish in a statistically isotropic, parity-symmetric theory, making them useful probes of parity-violating physics in the early Universe. A rotation of the polarization plane of $E$ modes into $B$ modes during propagation, usually called cosmic birefringence is what generates nonzero $TB$ and $EB$ correlations. This effect has been studied for example, in Chern-Simons modifications of electrodynamics and models with cosmological pseudoscalar fields~\cite{CarrollFieldJackiw1990,Carroll1998,LueEtAl1999}, as well as in the context of Lorentz and {\it CPT} violation, which can generate the same parity-odd operator and the same polarization signature~\cite{NilssonLePoncinLafitte2023}.
Encouragingly, hints of isotropic birefringence were reported using \textit{Planck} polarization data~\cite{MinamiKomatsu2020}, with subsequent analyses examining foreground emission and frequency dependence~\cite{DiegoPalazuelosEtAl2022,Eskilt2022}. Together with WMAP data, the constraint on the cosmic birefringence angle reads $\beta=(0.342^{+0.094}_{-0.091})^\circ$ at 68\% confidence~\cite{birefringence-2022}.
More recently, the Atacama Cosmology Telescope (ACT) give a rotation of comparable magnitude around $2\sigma$ confidence~\cite{DiegoPalazuelosKomatsu2025}, and joint ACT-\textit{Planck} analysis found $\beta=0.277^\circ\pm0.057^\circ$, with nominal significance $4.8\sigma$, reduced to $3.5\sigma$ in a test that reduces dust contamination~\cite{Eskilt2026}.

There are, however, several potential sources of error. In particular, isotropic birefringence is degenerate with the so-called miscalibration angle (normally called $\alpha$), a uniform error in the instrument polarization angle. With this in mind, it is known that calibration by imposing vanishing $TB$ and $EB$ can remove a physical signal~\cite{KeatingEtAl2013}, which risks biasing results. Using the different effects of rotation on CMB and Galactic emission (for example, Faraday rotation) can break this degeneracy~\cite{MinamiEtAl2019}, but this requires control of intrinsic foreground correlations. These observational difficulties and their implications are reviewed in ~\cite{Komatsu2022}. Axion-like particles provide an attractive way to generate such a rotation of $E$ modes into $B$ modes. The original axion follows from the Peccei-Quinn solution of the strong \emph{CP} problem~\cite{PecceiQuinn1977,Weinberg1978,Wilczek1978}, while more general light pseudoscalars occur in string constructions~\cite{SvrcekWitten2006,ArvanitakiEtAl2010} and can contribute to dark matter or dark energy depending on the relevant scales~\cite{Marsh2016}. Simultaneously, solar physics and stellar evolution can independently constrain their electromagnetic coupling, see for example~\cite{CAST2017,AyalaEtAl2014}. For the canonical axion-photon interaction $\phi F_{\mu\nu}\tilde F^{\mu\nu}$ and in the case of isotropic cosmic birefringence, the leading geometric-optics rotation is proportional to the axion field difference between emission and observation, which makes it sensitive to the cosmological background evolution.~\cite{FujitaEtAl2021}. On the other hand, spatial fluctuations in the axion field can generate anisotropic rotation~\cite{PospelovEtAl2009}, while rapid oscillations can produce polarization washout and time-dependent signals~\cite{FedderkeEtAl2019}, but this has not been observed to date.
 
There is, however, no reason for the axion to couple only to the visible photon. An
additional Abelian gauge field can interact with the visible sector through kinetic
mixing~\cite{Holdom1986}; reviews are given in Refs.~\cite{JaeckelRingwald2010,FabbrichesiEtAl2020}.
In this paper, we consider an unbroken dark U(1) whose massless dark photon forms a radiation
bath rather than dark matter~\cite{Ackerman:2008kmp}. We assume that any dark charged particles have negligible abundance at the epochs of interest. The dark photons then
free-stream and retain the thermal spectrum established while the dark sector was in
equilibrium, so there is no dark plasma and their temperature relative to the photons is a
free parameter constrained through $N_{\rm eff}$~\cite{Planck-2018}.
With both an axion and a dark photon present, the mixed interaction
$\phi F_{\mu\nu}\tilde{B}^{\mu\nu}$ and the dark interaction
$\phi B_{\mu\nu}\tilde{B}^{\mu\nu}$ are also allowed. Here $B_{\mu\nu}$ is the dark field
strength. These dark axion portal couplings need not be fixed by kinetic mixing~\cite{KanetaEtAl2017}.
Related studies consider axion-mediated mixing in external magnetic fields~\cite{Ejlli2018}. 
The same couplings can also drive nonperturbative production of dark gauge fields from
the axion background, with backreaction on the axion and changes in its abundance
~\cite{ChoiEtAl2017,AgrawalEtAl2018,KitajimaEtAl2018}. That process is distinct from the conversion of pre-existing radiation studied
here, but it bears on our assumption of a fixed axion background.

Some particularly relevant work already exists in the literature. \cite{DW-DP-2023} studied the transfer of dark photon birefringence to visible photons, including spectral distortion and circular polarization. Their axion example contains the general three-coupling system and its propagation eigenstates. Similarly, CMB spectral-distortion constraints from the mixed interaction when either the axion or dark photon supplies a dark matter background were derived in~\cite{HookEtAl2023}. In this work, we consider the combined polarization and intensity constraints when the dark photon is massless, both gauge sectors contain radiation, and all three axion couplings are present.

For the purposes of constraining the radiation conversion fraction, the CMB monopole spectrum is a particularly sensitive probe.  COBE/FIRAS established its close agreement with a blackbody~\cite{MatherEtAl1994,CMB-1996} at approximately $2.725\,\mathrm{K}$~\cite{Fixsen2009}. Spectral distortions closely depend on the process changing the radiation and the efficiency of thermalization~\cite{ChlubaSunyaev2012,Chluba2015}, and spectral constraints on kinetically mixed massive dark photons already exploit this sensitivity~\cite{MirizziEtAl2009,CaputoEtAl2020}, although their plasma-dependent conversion probabilities cannot be applied directly to the present model. Mixing initially thermal populations at different temperatures generally produces a non-blackbody spectrum~\cite{KhatriEtAl2012}, even for frequency-independent conversion. Its shape need not be a Compton $y$ distortion, and a comparison with observations must allow the fitted temperature to change. More sensitive spectroscopy, as discussed for PIXIE and related proposals, could further constrain such signals~\cite{KogutEtAl2011,ChlubaEtAl2019}.

In this work, we study the three couplings $\phi F\tilde F$, $\phi B\tilde F$, and $\phi B\tilde B$ for massless gauge fields in a homogeneous axion-like dark matter background on a spatially flat Friedmann-Lema\^itre-Robertson-Walker (FLRW) spacetime. Within the vacuum propagation and leading adiabatic approximations, we derive the field evolution, distinguish coherent initial conditions from phase-averaged thermal quantities, and compare the resulting spectrum and rotation with observations. The main question we ask is how the dark sector changes the inference of the ordinary axion-photon coupling and how radiation conversion restricts that freedom.
We introduce the model with axion-photon-dark photon couplings in Section~\ref{ch2}. In Section~\ref{ch3}, we present the propagation solutions of photon and dark photon fields and analyze their evolutions in several special cases. Section~\ref{sec:CMB} contains the observational constraints from CMB spectral distortion and cosmic birefringence, followed by our conclusions in Section~\ref{ch5}.

Throughout this work, we use natural units with $c=k_B=\hbar=1$ and use the metric signature $(-+++)$.  The scale factor is defined as $a(t)=1/(1+z)$ and normalized as $a(t_0)=1$, where $t_0$ denotes the present time. Accordingly, the comoving and physical wavenumbers are related by $k_c = a k_p$.

\section{The model} \label{ch2}

The electromagnetic field $A^\mu$ and dark photon field $B^\mu$ in the kinetic-diagonal basis couple to the axion field $\phi$ with the Lagrangian\footnote{The canonical normalization is shown in Appendix \ref{app-1}. Note that $B^{\mu\nu}\tilde{F}_{\mu\nu}=F^{\mu\nu}\tilde{B}_{\mu\nu}$, so only one of these equivalent terms needs to be included in Eq.~\eqref{eq-L}.}
\begin{equation}
 \cL = - {\frac12}\partial_{\mu}\phi\partial^{\mu }\phi- V(\phi) - \frac{1}{4}F_{\mu\nu}F^{\mu\nu} - \frac{1}{4}B_{\mu\nu} B^{\mu\nu} +g_{1}\phi F^{\mu\nu}\tilde{F}_{\mu\nu} + g_{2}\phi B^{\mu \nu }\tilde{F}_{\mu\nu}+g_{3}\phi B^{\mu\nu} \tilde{B}_{\mu\nu}, \label{eq-L}
\end{equation}
where the corresponding field strengths are
\begin{equation}
\begin{aligned}
    {F}_{\mu\nu}\equiv& \nabla_\mu {A}_\nu - \nabla_\nu {A}_\mu, \\
    {B}_{\mu\nu}\equiv& \nabla_\mu {B}_\nu - \nabla_\nu {B}_\mu,
\end{aligned}
\end{equation}
and their duals are defined through the Levi-Civita tensor as
\begin{equation}
    \quad \tilde{{F}}^{\mu\nu}\equiv \frac{1}{2}\varepsilon^{\mu\nu\rho\sigma}{F}_{\rho\sigma}, \quad \tilde{{B}}^{\mu\nu}\equiv \frac{1}{2}\varepsilon^{\mu\nu\rho\sigma}{B}_{\rho\sigma},
\end{equation}
where $\varepsilon^{\mu\nu\rho\sigma} \equiv \eta^{\mu\nu\rho\sigma}/\sqrt{-g}$ is the Levi-Civita tensor.

By varying the Lagrangian, we obtain the equations of motion for $A^\mu$ and $B^{\mu}$ as
\begin{align}
\label{eq:Aeom}\nabla_\mu F^{\mu\nu}
-4 g_1\,(\nabla_\mu\phi)\,\tilde F^{\mu\nu}
-2 g_2\,(\nabla_\mu\phi)\,\tilde B^{\mu\nu}
&=0, \\[6pt]
\label{eq:Beom}\nabla_\mu B^{\mu\nu}
-4 g_3\,(\nabla_\mu\phi)\,\tilde B^{\mu\nu}
-2 g_2\,(\nabla_\mu\phi)\,\tilde F^{\mu\nu}
&=0.
\end{align}

Here we adopt the Coulomb gauge $\nabla\cdot \mbf{A} = \nabla\cdot \mbf{B} = 0$, and for electromagnetic and dark photon fields in vacuum, the scalar potentials vanish, i.e. $A^0 = B^0 = 0$.
The equations of motion for $A^\mu$ and $B^\mu$ are
\begin{align}
&\partial_\mu\partial^\mu \mbf{A} - 4g_1a\dot\phi\,\nabla\times\mbf{A} - 2g_2a\dot\phi\,\nabla\times\mbf{B} \ = \ 0 \label{EOM-1-1},\\
&\partial_\mu\partial^\mu \mbf{B} - 4g_3a\dot\phi\,\nabla\times\mbf{B} - 2g_2a\dot\phi\,\nabla\times\mbf{A} \ = \ 0 \label{EOM-1-2}.
\end{align}

Let us consider the helicity modes of the potential fields, $\mbf{A}(\mbf{x},\eta) = \hat{\mbb\epsilon}_\pm A_\pm(\mbf{k},\eta)e^{i\mbf{k}\cdot\mbf{x}}$ and $\mbf{B}(\mbf{x},\eta) = \hat{\mbb\epsilon}_\pm B_\pm(\mbf{k},\eta)e^{i\mbf{k}\cdot\mbf{x}}$, where helicity eigenvectors $\hat{\mbb\epsilon}_\pm$ have property $i\mbf{k}\times \hat{\mbb\epsilon}_\pm = \pm k\hat{\mbb\epsilon}_\pm$ so we have
\be
\nabla\times\mbf{A} = \pm k\, \hat{\mbb\epsilon}_\pm A_\pm(\mbf{k},t)e^{i\mbf{k}\cdot\mbf{x}}\ ,\ \nabla\times\mbf{B} = \pm k\, \hat{\mbb\epsilon}_\pm B_\pm(\mbf{k},t)e^{i\mbf{k}\cdot\mbf{x}}.
\ee
Inserting the FLRW metric, the equations of motion in Fourier space become
\begin{align}
    A_\pm'' + k_c^2 A_\pm \pm 4g_1 \phi' k_c A_\pm \pm 2g_2 \phi' k_c B_\pm = 0 \label{EOM-2-1},\\
    B_\pm'' + k_c^2 B_\pm \pm 4g_3 \phi' k_c B_\pm \pm 2g_2 \phi' k_c A_\pm = 0  \label{EOM-2-2}.
\end{align}
where $'$ denotes the derivative w.r.t. the conformal time $\eta$ ($d\eta= dt/a\ \Rightarrow\ \eta = 3t_0^{2/3}t^{1/3}$ for matter-dominated era), and $k_c$ is the comoving wavenumber.

The equation of motion for the scalar field $\phi$ is
\begin{align}
    &\Box\phi
    -V_{,\phi}+g_{1}\,F^{\mu\nu}\,\tilde{F}_{\mu\nu} +g_{2}\,B^{\mu\nu}\,\tilde{F}_{\mu\nu}+g_{3}\,B^{\mu\nu}\,\tilde{B}_{\mu\nu}=0 .
\end{align}
Ignoring the backreaction due to the axion-photon (dark photon) coupling\footnote{Due to the tight constraint on the coupling constant, the influence of the radiation and dark radiation fields on the axion field is negligible without considering parametric resonance.}, the EOM becomes
\begin{align}
    \ddot\phi + 3H\dot\phi + m^2\phi = 0,
\end{align}
and the scalar field follows an oscillating solution
\be
    \phi(t)=\phi_0(t) \sin(mt),
\ee
where the energy density of the axion field is $\rho_\phi= m^2\phi_0^2/2$. 
In an expanding universe, the amplitude of the axion field $\phi_0$ should also evolve over time\footnote{Note that we are interested in the impact of the axion-photon-dark photon coupling on CMB photons, so we only consider the evolution during the matter-dominated era.} as
\begin{align}
    \phi_0(t) &= \frac{1}{m}\sqrt{2f_{\phi}\rho_{m}} = \phi_{rec}\Big(\frac{z+1}{z_{rec}+1}\Big)^{3/2}, 
\end{align}
where $f_\phi\equiv \rho_\phi/\rho_{m}$ is the axion abundance fraction, and $\rho_{m}$ is the matter energy density in the universe.

\section{Cosmic birefringence and energy conversion from axion-photon-dark photon coupling}\label{ch3}

In this section, we solve the above EOM for the gauge field and study the cosmic birefringence and energy conversion due to the axion-photon-dark photon coupling.


If we rewrite the equations in eq.\eqref{EOM-2-1} and \eqref{EOM-2-2} in a matrix notation
\be
\vec{A}_\pm = \left(\begin{array}{c}
    A_\pm\\
    B_\pm 
\end{array}\right),\ \ 
C = \left(\begin{array}{cc}
    4g_1 & 2g_2 \\
    2g_2 & 4g_3
\end{array}\right),
\ee
and the EOM becomes
\begin{align}
L \vec{A}_\pm &= \pm(\phi'k_c) C \vec{A}_\pm,
\end{align}
where the operator is $L\equiv -\partial_{\eta}^2 - k_c^2$. As a real symmetric matrix is diagonalizable  we have that $C=MDM^{-1}$ where $D=\text{diag}(\lambda_1,\lambda_2)$ is a diagonal matrix and $M$ is an orthogonal matrix defined in Appendix~\ref{app-2}. The two eigenvalues of $D$ are
\begin{align}
\begin{array}{l}
    \lambda_1 = 2\big( g_1 + g_3 + \sqrt{(g_1-g_3)^2 + g_2^2} \big), \\
    \lambda_2 = 2\big( g_1 + g_3 - \sqrt{(g_1-g_3)^2 + g_2^2} \big).
\end{array}
\end{align}

Then, for the propagation eigenstates $\vec{\mcal{A}}_\pm = M^{-1} \vec{A}_\pm \equiv (\mcal{A}_1^\pm, \mcal{A}_2^\pm,)^T$, we have the diagonal EOM $L\vec{\mcal{A}}_\pm = \pm(\phi' k_c) D \vec{\mcal{A}}_\pm$, i.e.
\begin{align}
    \partial_\eta^2\mcal{A}_b^\pm + k_c^2\mcal{A}_b^\pm \pm \lambda_b \phi' k_c \mcal{A}_b^\pm = 0
\end{align}
for $\mcal{A}_b^\pm$ ($b=1,\ 2$).

In the expanding universe, the plane wave factor of an electromagnetic wave with a fixed comoving wavenumber $k_c$ is
\begin{align}
    \mcal{A} &= \mcal{A}_0 \exp\Big[ i\,\mbf{k}_c\cdot\mbf{x}_c - i\int\omega_b(\eta) \,d\eta \Big].
\end{align}
Thus, the propagation eigenstates in Fourier space evolve as $\mcal{A}_b^\pm =  \mcal{A}_{0,b}^\pm e^{-i\int\omega_b^\pm d\eta}$. Then, we can derive 
\begin{align}
    (\omega_b^\pm)^2 = k_{c}^2 \pm \lambda_b \phi' k_{c}.
\end{align}

Since the axion-photon-dark photon coupling terms are significantly weaker compared to the $k_c^2$ term\footnote{If we consider the overall upper bound on the axion-photon coupling constant $g_1\lesssim 10^{-10}\,{\rm GeV}^{-1}$, we have $g_1 \phi'\sim g_1 a m\phi(\eta)\lesssim 10^{-23}{\rm eV}$ at recombination and it will decrease over time. This value is $\sim20$ orders of magnitude smaller than the comoving wavenumber of interest. Therefore, this approximation is valid if the other coupling constants are not many orders of magnitude larger than $g_1$.}, i.e. $\lambda_b\phi' k_c \ll k_c^2$, we have
\be
    \omega_b^\pm \approx k_c \pm \frac12 \lambda_b \phi',
\ee
and the final time-dependent phase of this wave becomes
$\Phi_{\cA_b^\pm} = -\int \omega_b^\pm(\eta)d\eta \,\approx\, -\big(k_c\cdot \eta \pm \frac{1}{2}\lambda_b \int \phi' d\eta\big).
$
To simplify expressions involving integration over $\phi'$, we define
\begin{equation}\label{eq:Iphi_def}
I_{\phi}(\eta)\equiv \int_{\eta_i}^{\eta} \phi'(\eta)d\eta = \phi(\eta)-\phi(\eta_i),
\end{equation}
which we use from now on.
Note that although the amplitude of the axion field $\phi_0$ decreases as $t^{-1}$, the value of $\phi$ at $\eta_i$ and $\eta$ also depends on the phase term and we do not know whether $I_{\phi}(\eta)$ is positive or negative. With this in mind, we can express $A_\pm$ and $B_\pm$ by $\mcal{A}_b^\pm$, assuming
\begin{align}
    A_\pm = \cos\alpha_1 \mcal{A}_1^\pm + \cos\alpha_2 \mcal{A}_2^\pm,\quad 
    B_\pm = \sin\alpha_1 \mcal{A}_1^\pm + \sin\alpha_2 \mcal{A}_2^\pm.
\end{align}
Note that there is only one free parameter in the orthogonal matrix $M$ but we use $\alpha_1$ and $\alpha_2$ to distinguish the contribution from the two eigenstates ${\cal A}_1$ and ${\cal A}_2$.

Inserting the time evolution of $\mcal{A}_b^\pm$ and the initial amplitude $\mcal{A}_{b,0}^\pm = \cos\alpha_b A_0^\pm + \sin\alpha_b B_0^\pm$
\begin{align}
    A_\pm &= \cos\alpha_1 (\cos\alpha_1 A_0^\pm + \sin\alpha_1 B_0^\pm) e^{-i\int\omega_1^\pm d\eta} + \cos\alpha_2 (\cos\alpha_2 A_0^\pm + \sin\alpha_2 B_0^\pm) e^{-i\int\omega_2^\pm d\eta} , \\
    B_\pm &= \sin\alpha_1 (\cos\alpha_1 A_0^\pm + \sin\alpha_1 B_0^\pm) e^{-i\int\omega_1^\pm d\eta} + \sin\alpha_2 (\cos\alpha_2 A_0^\pm + \sin\alpha_2 B_0^\pm) e^{-i\int\omega_2^\pm d\eta} .
\end{align}

Let us define a new set of coupling constants 
\begin{align}\label{eq:g0g4thg}
\left\{\begin{array}{l}
    g_0 \equiv g_1+g_3 , \\
    g_4 \equiv \sqrt{(g_3-g_1)^2+g_2^2} , \\
    \tan\thg \equiv \frac{g_2}{g_3-g_1} ,
\end{array}\right.
\end{align}
with $\thg\in [0,\pi]$. 
The new coupling constants will significantly simplify the results. The first parameter $g_0$ corresponds to the sum of $g_1$ and $g_3$, while the last two parameters $\{g_4,\,\thg\}$ correspond to the radial and angular coordinates in the $g_3-g_2$ plane with the origin located at $(g_1,0)$.
Note that they are a complete set of coupling constants, and we have
\begin{align}\label{eq:gthrelation}
\left\{\begin{array}{l}
    g_1 = \frac{1}{2} (g_0 - g_4\cos\thg) , \\
    g_2 = g_4 \sin\thg ,\\
    g_3 = \frac{1}{2} (g_0 + g_4\cos\thg).
    \end{array}\right.
\end{align}

With the new coupling constants, the two eigenvalues become
\begin{align}
    \lambda_{1,2} = 2 (g_0 \pm g_4),
\end{align}
where the subscripts $1$ and $2$ correspond to the $+$ and $-$ signs, respectively.

Additionally, we can find that $\alpha_1$ and $\alpha_2$ are completely determined by $\thg$\footnote{The details of the calculation are shown in Appendix~\ref{app-2}.}
\begin{align}
    &\alpha_1 = \frac{3}{2}\pi-\frac{\thg}{2},\quad\alpha_2 = \pi-\frac{\thg}{2}.
\end{align}

$\{g_0,\ g_4,\ \thg\}$ (or $\{g_1,\ g_2,\ g_3\}$) is the complete set of free parameters related to the axion-photon-dark photon couplings. To obtain the evolution of photon and dark photon fields, we also need to know the parameters related to the initial conditions.


Note that only photon and dark photon modes with the same comoving wavevector $\mbf{k}_c$ are coupled in Fourier space. From now on, we consider linearly polarized initial radiation fields with the same momentum, $A_0^\pm(k_c)$ and $B_0^\pm(k_c)$. There are three free parameters related to the initial condition: the modulus of the ratio of their amplitude $\tR_0\equiv |B_0|/|A_0|$, the initial relative phase between dark photon and photon $\tPhi$, and the initial relative angle between the two polarization planes $\tphi$ (i.e. the relative angle between $A_0^\pm$ and $B_0^\pm$ in the complex plane).

In the helicity basis, we have the relation\footnote{The details of the calculation are shown in Appendix~\ref{app-3}.}.
\begin{align}\label{eq:BAamplitude_def}
    B_0^{\pm}/A_0^{\pm} &\equiv R_0^\pm=  \tilde{R}_0 e^{i(\tPhi\pm\tphi)}.
\end{align}
We can find that both $\tPhi$ and $\tphi$ act as phase terms of the initial fields. For simplification, we define the effective initial phase $\tphi^\pm\equiv \tPhi\pm\tphi$.

Then, we can express the two gauge fields as slowly modulated waves
\begin{equation}
\boxed{
\begin{aligned}
    A_\pm &= A_0^\pm R_A^\pm(\eta) \exp\big[-i\big(k_c\cdot\eta\pm g_0I_\phi(\eta) + \varphi_A^\pm(\eta)\big)\big] , \\
    B_\pm &= B_0^\pm R_B^\pm(\eta) \exp\big[-i\big(k_c\cdot\eta\pm g_0I_\phi(\eta) + \varphi_B^\pm(\eta)\big)\big] .
\end{aligned}} \label{eq-3-17}
\end{equation}
where 
\begin{align}
    R_A^\pm(\eta)^2 
    &= 1 - (1-\tR_0^2)\sin(\thg)^2 \sin\big(g_4 I_\phi(\eta)\big)^2 \nn
    &\quad\quad+ 2 \tR_0\Big[ \pm \sin(\tphi^\pm)\cos\big(g_4 I_\phi(\eta)\big) - \cos(\tphi^\pm)\cos(\thg)\sin\big(g_4 I_\phi(\eta)\big) \Big] \sin(\theta_g)\sin\big(g_4 I_\phi(\eta)\big)\label{eq-RA} , \\
    R_B^\pm(\eta)^2 
    &= 1 - (1-\tR_0^{-2})\sin(\thg)^2 \sin\big(g_4 I_\phi(\eta)\big)^2 \nn
    &\quad\quad+ 2 \tR_0^{-1}\Big[ \mp \sin(\tphi^\pm)\cos\big(g_4 I_\phi(\eta)\big) + \cos(\tphi^\pm)\cos(\thg)\sin\big(g_4 I_\phi(\eta)\big) \Big]\sin(\theta_g)\sin\big(g_4 I_\phi(\eta)\big) \label{eq-RB} ,
\end{align}
and 
\begin{align}
    \label{eq:phipmA}\varphi_A^\pm &= \mp \arctan\Bigg[\frac{\big(\cos(\theta_g) -\tR_0\cos(\tphi^\pm)\sin(\theta_g)\big)\tan\big(g_4 I_\phi(\eta)\big)}{1 \pm \tR_0\sin(\tphi^\pm)\sin(\theta_g)\tan\big(g_4 I_\phi(\eta)\big)}\Bigg] , \\
   \varphi_B^\pm &= \pm\arctan\Bigg[\frac{\big(\cos(\theta_g) + \tR_0^{-1}\cos(\tphi^\pm)\sin(\theta_g)\big)\tan\big(g_4 I_\phi(\eta)\big)} {1 \mp \tR_0^{-1}\sin(\tphi^\pm)\sin(\theta_g)\tan\big(g_4 I_\phi(\eta)\big)}\Bigg] . \label{eq-3-21}
\end{align}
Note that $k_c\gg g_i\phi'$, so $g_4 I_\phi(\eta)$ varies significantly more slowly than the $k_c\cdot \eta$, which means the $k_c\cdot\eta$ term characterizes the dominant oscillatory behaviour of the wave. Thus, the amplitude evolution factors $R_X^\pm$, and the rest phase terms $\pm g_0I_\phi(\eta) + \varphi_X^\pm(\eta)$ for the photon or dark photon field $X=A,\ B$, can be viewed as slowly varying quantities.
The amplitude factors $R_X^{\pm}$ describe the change in the field amplitudes due to energy conversion between photons and dark photons, while the cosmic birefringence angles for the $X=A, B$ fields are related to the slowly varying phase as\footnote{As the two helicity modes propagate independently, the birefringence angles derived from the Stokes parameters are the difference between their phases. The proof is shown in App.~\ref{app-4}.}
\begin{equation}\label{eq:betaX_def}
\begin{aligned}
\beta_X & = -\frac12 \arg\Big(\frac{X_+ X_-^*}{X_0^+X_0^{-*}}\Big)= -\frac12 (\Phi_X^+-\Phi_X^-) \\&= g_0 I_{\phi}(\eta) + \frac12(\varphi_X^+ - \varphi_X^-).
\end{aligned}
\end{equation}
They are able to influence the CMB signals and therefore can be constrained by CMB observations.

To summarize, seven parameters appear in the results, which are as follows 
\bi

    \item the three coupling constants related to the interactions between the related to axion-photon-dark photon\footnote{Note that any three of these paramters, except for $\{g_1,\ g_3,\ g_0\}$, constitute a complete parameter set for coupling constants. They are related as shown in Eq.~\eqref{eq:gthrelation}.}
    $\{g_1,\ g_2,\ g_3\}$, as defined in Eq.~\eqref{eq-L}
    or equivalently $\{ g_0,\ g_4,\ \thg \}$, as defined in Eq.~\eqref{eq:g0g4thg};
    
    \item the change in the axion field $I_\phi = \int \phi'(\eta)d\eta =\phi(\eta)-\phi(\eta_i)$ defined in Eq.~\eqref{eq:Iphi_def}.
    However, we note that $I_\phi$ is degenerate with an overall rescaling of the coupling constants $g_i$, i.e. these expressions can be entirely rewritten in terms of the combinations $g_i I_\phi$.
    Thus, we define the dimensionless parameters 
    \be\label{eq:betai_def}
        \beta_i=2g_i I_\phi,
    \ee
    where $\beta_1$ is the canonical cosmic birefringence angle (i.e. coming from the $\phi F\tilde{F}$-term only), and the time dependence of $R_X^\pm(\eta)$ and $\varphi_X^\pm(\eta)$ enters only through $\beta_4(\eta)=2g_4 I_\phi(\eta)$;
    
    \item the three free parameters related to the initial condition of the photon and dark photon fields $\{ \tR_0,\ \tphi^\pm= \tilde{\Phi}\pm \tphi \}$ defined in Eq~\eqref{eq:BAamplitude_def}, 
    where $\tR_0=|B_0|/|A_0|$ is the modulus of the ratio between the initial dark photon amplitude and the initial photon amplitude, $\tilde{\Phi}$ is the initial relative phase between dark photon and photon, and $\tphi$ is the initial relative angle between the two polarization planes\footnote{Note that when $\tPhi=0$, i.e., $\tphi^-=-\tphi^+$, or when averaging over $\tPhi$, we have $R_X^+=R_X^-$ and $\varphi_X^-=-\varphi_X^+$. Given the tight observational constraints on the circular polarization of CMB photons, it is reasonable to neglect the initial relative phase $\tPhi$. In this case, we have
    $$
        \beta_X = g_0 I_\phi + \varphi_X^+,
    $$
    so we focus on the ``$+$" helicity mode in the following.}. In the helicity basis, we have $B_0^\pm/A_0^\pm = \tR_0 e^{i\tphi^\pm}$. 
    
\ei

\begin{figure}
    \centering
    \includegraphics[width=0.48\linewidth]{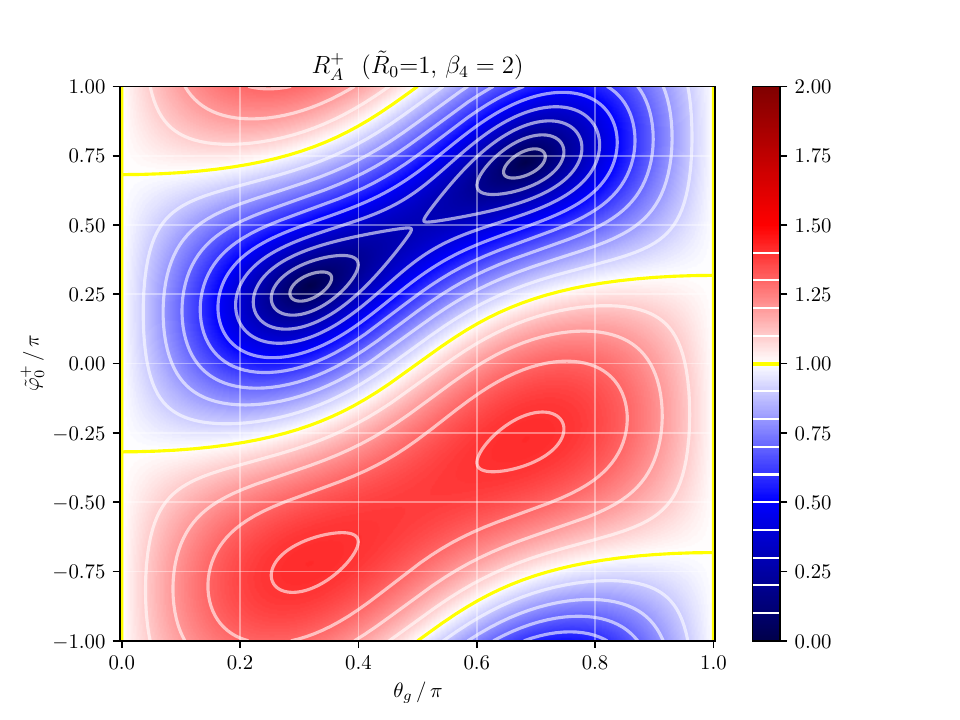}
    \includegraphics[width=0.48\linewidth]{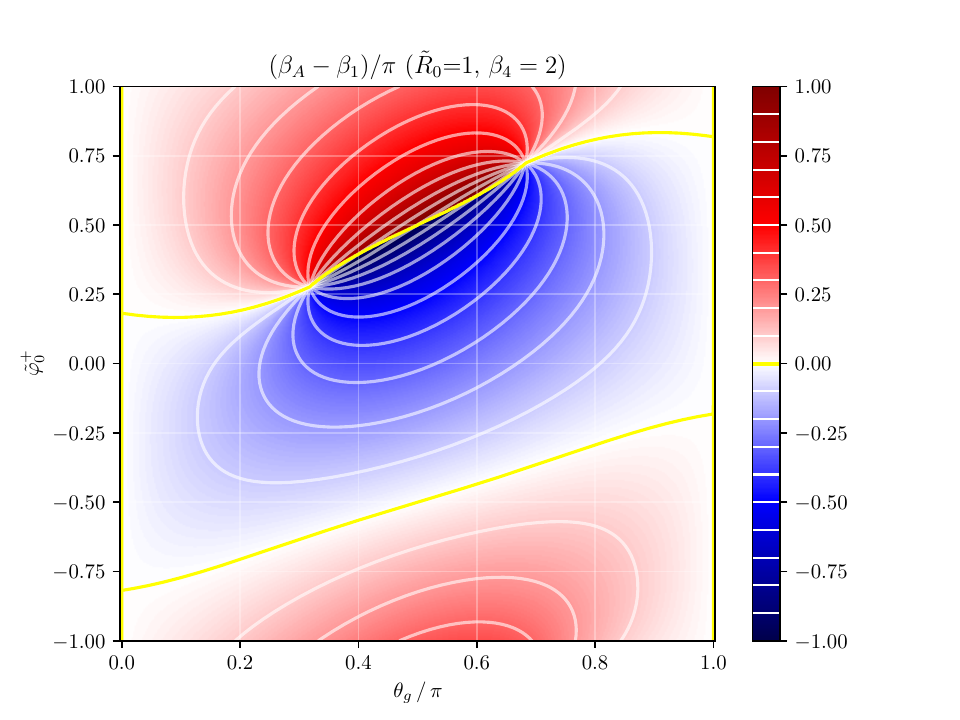}
    \caption{\tit{Left panel}: The photon amplitude ratio $R_A^\pm(t_0)$ in Eq.~\eqref{eq-RA} as a function of $\theta_g$ and $\tilde\varphi$. The solid white curves are contour lines with a contour interval of 0.1, whereas the yellow curves correspond to $R_A^+=1$. \tit{Right panel}: The difference between the $A$ field birefringence angle $\beta_A$ and $\beta_1=2 g_1 I_\phi$. The solid white curves are contour lines with a contour interval of $0.1\pi$, while the yellow curves correspond to $\beta_A=\beta_1$.}
    \label{fig1}
\end{figure}
In Figure~\ref{fig1}, we show the photon amplitude ratio $R_A^\pm(t_0)$ and $\Delta\beta_A\equiv \beta_A-\beta_1$ as functions of $\theta_g$ and $\tphi^+$ for initial amplitude ratio $\tilde{R}_0=1$ and $\beta_4=2$, i.e. $g_4I_\phi=1$. The left panel shows periodic behaviour with islands of $R^+_A$ varying from $0$ to $\sqrt2$, i.e. $(R^+_A)^2\in [0,2]$. Each island has two maximal/minima both located at around $\theta_g/\pi\approx 0.3$ and $0.7$. Reducing $\tilde{R}_0$ whilst keeping $\beta_4$ constant makes lower values of $R_A^\pm(t_0)$ dominate the parameter space, and increasing $\tilde{R}_0$ causes the opposite to happen. On the other hand, keeping $\tilde{R}_0$ fixed whilst reducing $\beta_4$ merges the two islands; for example, for $\tilde{R}_0=\beta_4=1$, $R^+_A$ exhibits one region with higher values and one with lower, bounded by a single $R^+_A=1$ line around $\tilde{\varphi}^+_0/\pi\in\,(-1/4,+1/4)$. In a similar but opposite way, increasing $\beta_4$ increases the amount of high and low $R^+_A$ islands in parameter space. The right panel similarly shows $\Delta\beta_A$, the difference between the total birefringence angle and that of the standard axion-photon case, which shows similar periodic behaviour. In this case, reducing the initial amplitude ratio $\tilde{R}_0$ significantly moves the birefringence angle towards the standard case of $\beta_1$ throughout parameter space; conversely, increasing the value of $\tilde{R}_0$ shows increased oscillations in $\Delta\beta_A$ as the coupling constants change.

\subsection{Special cases}

To study the properties of the model in more detail, we consider several special cases in parameter space. Note that since we can only observe the photon field directly, we restrict our attention to the behaviour of $A_\pm$ in this section.

\subsubsection{Incoherent case} \label{ch-case0}

Note that the initial relative phase and angle between $A_0$ and $B_0$ are well-defined only when both the initial photon and dark photon fields are coherent. However, if we are interested in large-scale signals, such as the spectral distortion of the CMB monopole intensity, we need to use the spatially averaged result. It is reasonable to assume that the initial relative phase $\tPhi$ and relative angle $\tphi$ are randomly distributed in a region much larger than the coherence volume, while all other parameters are spatially uniform. Thus, the spatial average can be reduced to an average over the two initial phase parameters $\tphi$ and $\tPhi$.

For a vector field $\mathbf{X}$, its spatial average  $\langle\mathbf{X}\rangle$ can vanish when the field carries a finite energy density, due to its random direction. Therefore, the radiation field should not be averaged at the field-amplitude level. Instead, we consider the spatially averaged radiation intensity, which is proportional to $\langle I_X \rangle =  k^2|\mathbf{X}|^2\propto (R_X^\pm)^2$.

\begin{figure}
    \centering
    \includegraphics[width=0.45\linewidth]{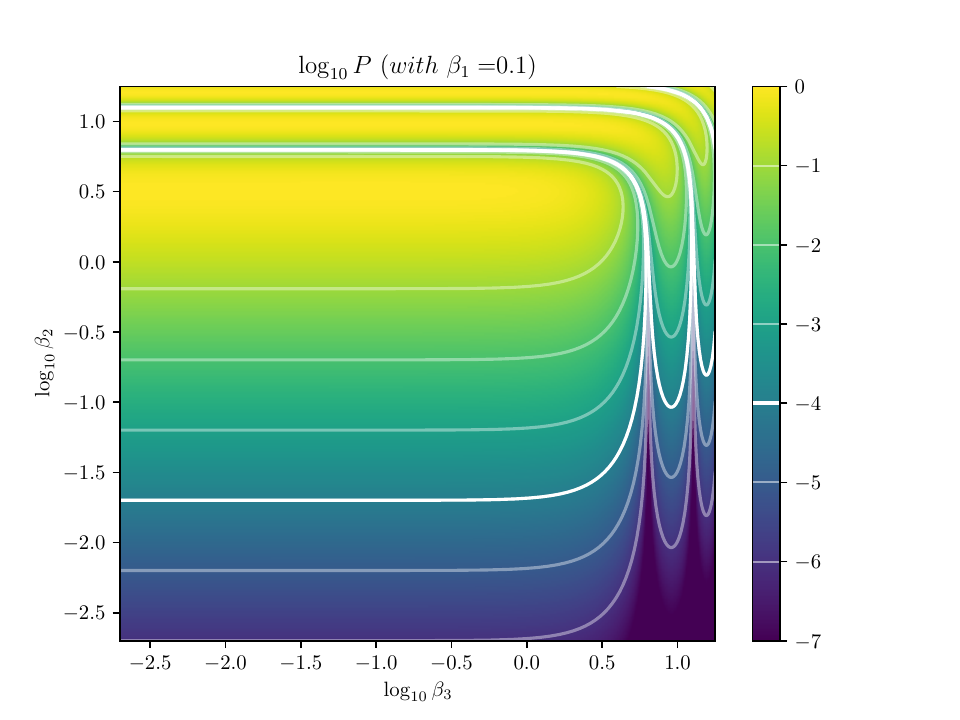}
    \caption{Contour plot of $P$ as a function of $g_3 I_\phi(\eta)$ and $g_2 I_\phi(\eta)$ for $g_1 I_\phi(\eta)=0.1$. Since $g_1$ and $g_3$ appear symmetrically in the expression for $P$, the dependence of $P$ on $g_1$ can be equivalently yielded from this figure by interchanging $g_1$ and $g_3$. The white contour lines correspond to $\log_{10}P=-1,-2,\dots,-6$ separately.}
    \label{fig2}
\end{figure}

Averaged over the initial phase $\tphi^\pm$, only the first two terms in eqs.~\eqref{eq-RA} and \eqref{eq-RB} preserve, and we have
\begin{align}
    \langle R_A^\pm{}^2 \rangle &= 1 - (1-\tR_0^2)\sin(\thg)^2 \sin\big(\beta_4/2\big)^2, \label{eq-3-24}\\
    \langle R_B^\pm{}^2 \rangle &= 1 - (1-\tR_0^{-2})\sin(\thg)^2 \sin\big(\beta_4/2\big)^2. \label{eq-3-25}
\end{align}
After the averaging, the amplitudes of the two helicity modes are exactly the same, indicating no global circular polarization due to this mechanism.
From these expressions, we can define the fraction of the global energy conversion as
\begin{align}\label{eq:P_def}
    P\equiv \sin(\thg)^2 \sin\big(\beta_4/2\big)^2 = \Big[\frac{\beta_2}{\beta_4} \sin\big(\beta_4/2\big) \Big]^2,
\end{align}
with
\begin{align}
    \left(\begin{array}{c}
        \rho_A \\
        \rho_B
    \end{array}\right) = 
    \left(\begin{array}{cc}
        1-P & P \\
        P & 1-P
    \end{array}\right)\left(\begin{array}{c}
        \rho_{A,0} \\
        \rho_{B,0}
    \end{array}\right),\label{eq-3-28}
\end{align}
where $\rho_{X,0}$ is the initial energy density of the $X$ field. Note that $P$ only depends on the axion-photon-dark photon interaction and is independent of the frequency $k_c$ or the radiation intensity. Figure \ref{fig2} shows the value of $P$ as a function of $g_i$.

For small $\beta_4$, the conversion fraction is approximately
\begin{align}
    P \sim \beta_2^2/4,
\end{align}
which depends only on the mixing coupling constant $g_2$. 
However, with the increase of $|g_3-g_1|$, the approximation $\sin(\beta_4/2)\sim \beta_4/2$ is no longer valid. This leads to the peaks in the contour lines on the right-hand side of the figures, where $P$ becomes small for the same value of $\beta_2$, due to the increasing deviation of $\sin(\beta_4/2)$ from $\beta_4/2$. In this region, the constraint on $\beta_2$ from the energy conversion fraction can be relaxed.
The thick white curves are the contour lines with $P=10^{-4}$, which roughly corresponds to the upper bound of the energy conversion fraction from CMB spectral distortion. As such, the parameter space \emph{above} this solid white line should be considered ruled out (see Section~\ref{sec:CMB} for further details).

Note that in Fig.~\ref{fig2} there are two narrow regions around $\beta_2(\eta_0)=2\pi$, $4\pi$ (the top two white bands)  where the value of $P$ decreases significantly, because $\sin(\beta_4(\eta_0)/2)\rightarrow 0$ in these regions. However, this does not imply that the photon and dark photon fields have experienced very weak conversion during their evolution. For the parameter values in these narrow regions, the field amplitudes change significantly $\Delta R_A^\pm|_{max} \sim (1-\tR_0^2)\frac{g_2^2}{g_4^2}$ at earlier times but return to values close to their initial values due to the periodic property\footnote{Although $R_X^\pm$ is not a periodic function of the cosmic time, it is a periodic function of the time-dependent parameter $\beta_4(\eta)=2g_4 I_\phi(\eta) = 2 g_4\int \phi' d\eta$, and it satisfies $R_X^\pm|_{\beta_4=2n\pi} = R_X^\pm|_{\beta_4=0} = 1$.} of $R_X^\pm$ as $I_\phi(\eta)$ evolves from $\eta_i$ to $\eta_0$ (the present time). 
In these cases, $R_X^\pm(\eta_0)$ is close to $1$ at the present time because $I(\eta_0)$ happens to be around $n\pi/g_2$ and the small value of $P$ at the present time cannot reflect the “cumulative” energy conversion fraction. 
Therefore, we regard these regions as fine-tuned and do not consider them to be viable allowed parameter space.

The spatially-averaged cosmic birefringence angles for $A$ and $B$ fields becomes (detailed calculation see App.~\ref{app-4})
\begin{align}
    \langle\beta_A\rangle &= \frac{1}{2}(\beta_1+\beta_3) -  \arctan\big[\tfrac{\beta_3-\beta_1}{\beta_4}\tan\big(\beta_4/2\big)\big] , \label{eq-3-29}\\
    \langle\beta_B\rangle &= \frac{1}{2}(\beta_1+\beta_3) +  \arctan\big[\tfrac{\beta_3-\beta_1}{\beta_4}\tan\big(\beta_4/2\big)\big].\label{eq-3-30}
\end{align}
We can find that these angles are completely independent of the initial amplitudes of the two radiation fields and are determined by the coupling between them and the axion field. This is valid for a region much larger than the coherent scale of the photon and dark photon fields.

For the linear order of $\beta_4$, the two cosmic birefringence angles are decoupled, which only depend on their own couplings with $\phi$
\begin{equation}
    \langle\beta_X\rangle = 2 g_{\phi XX} I_\phi(\eta),
\end{equation}
where $g_{\phi AA}=g_1$ and $g_{\phi BB}=g_3$ are the coupling constant for the $\phi X\tilde{X}$ interation in our definition.

\subsubsection{Nonzero Initial Circular Polarization} \label{ch-case2}

In this subsection, we consider the case in which the initial radiation contains both linear and circular polarization components, i.e., $|X_0^+|\neq |X_0^-|$. Accordingly, the initial-condition parameters are generalized to
\begin{align}
B_0^\pm/A_0^\pm
= \tR_0^\pm e^{i\tphi^\pm},
\end{align}
where $\tR_0^\pm$ are positive real parameters with $\tR_0^+\neq \tR_0^-$. Since the two helicity modes evolve independently, we can follow the same procedure as in Appendix~\ref{app-3}, yielding the results in eqs.~ (\ref{eq-3-17}--\ref{eq-3-21}) with $\tR_0$ replaced by $\tR_0^\pm$.
Note that, although we continue to use $\tphi^\pm$ to denote the effective relative phases between $B_0^\pm$ and $A_0^\pm$, these phases are not simply given by the sum of $\tPhi$ and $\pm\tphi$. Instead, they also depend on the nonzero relative phase between $X_x$ and $X_y$. Nevertheless, on scales much larger than the coherence scale of the radiation fields, it remains reasonable to treat $\tphi^\pm$ as random variables and reduce the spatial average to the average over $\tphi^\pm$. Therefore, we can still use the expression in eqs.~(\ref{eq-3-24}--\ref{eq-3-25}) and (\ref{eq-3-29}--\ref{eq-3-30}) for global signals.

These results lead to an important conclusion: even if the initial photons are purely linearly polarized, their interaction with dark photons carrying a nonzero circular polarization component can generate circular polarization in the observed radiation field. 
Thus, observational constraints on the circular polarization of the CMB can also be used to constrain the circularly polarized component of the dark photon field. However, similar to the case of photon–dark photon energy conversion, only a small fraction of the circularly polarized dark photon modes $B_0^\pm$ is converted into the photon modes $A_\pm$:
\begin{align}
    V_A &= |A_0^+|^2\langle R_A^+{}^2\rangle - |A_0^+|^2\langle R_A^-{}^2\rangle = |A_0|^2\big[(\tR_0^+)^2-(\tR_0^-)^2\big]P = V_{B,0}P.
\end{align}

Different initial amplitude ratios $\tR_0^\pm$ will also influence the value of the local birefringence angle. 
However, if we integrate over the random initial phase $\tphi^\pm$, we obtain the same result as in the last subsection, which is independent of the initial conditions of the two fields.

\subsubsection{Weak parity-odd mixing} \label{ch-case1}

In the previous subsections, we showed the effects of special initial conditions, i.e. random initial relative phase and angle but with a fixed amplitude. In the following, we discuss some special cases of the coupling constants with coherent initial conditions.

First, we consider the special case with weak parity-odd mixing $|g_2|\ll|g_3-g_1|$, i.e. $\sin\thg\rightarrow 0$. 
Note that this condition is more restrictive than $g_2\ll g_1,g_3$ as it excludes the special case where $g_1 \approx g_3$, which may cause a relatively large $\sin\thg$.
The extreme case is $g_2=0$, which corresponds to the absence of the mixing interaction $\phi F\tilde{B}$. In this case, the result reduces to the standard axion-photon coupling, and we obtain\footnote{In this and the following subsection, we only show the results for the photon field $A_\pm$, as the dark photon field cannot be directly detected.}
\begin{align}
    &R_A^\pm = 1,\ \ \varphi_A^\pm = \pm\tfrac{1}{2}(\beta_1-\beta_3), \\
    \Rightarrow\ \ &A_\pm = A_0^\pm e^{-i[k_c\eta \pm \beta_1(\eta)]},\ \text{i.e. }\beta_A = \beta_1.
\end{align}

Then, considering a small value of $g_2$ compared to $|g_3-g_1|$, which corresponds to $\thg\rightarrow0$ or $\pi$, we can make an approximation keeping only linear order terms in $\sin\thg\simeq \beta_2/\beta_4$ that\footnote{Note that $\sin\thg\rightarrow 0$ corresponds to two cases $\sin\thg\sim\thg$ when $\thg\rightarrow0$ and $\sin\thg\sim \pi-\thg$ for $\thg\rightarrow\pi$. Thus, we use $g_2/g_4$ instead of $\thg$ to denote the small parameter.}, and we have
\begin{align}
    R_A^\pm &\simeq \left\{\begin{array}{ll}
        1 + \big[\pm\Im(R_0^\pm) \cos(\beta_4/2) -  \Re(R_0^\pm)\sin(\beta_4/2)\big]\sin(\beta_4/2)\tfrac{\beta_2}{\beta_4}, & \thg\rightarrow 0 , \\
        1 + \big[\pm\Im(R_0^\pm)\cos(\beta_4/2) + \Re(R_0^\pm)\sin(\beta_4/2)\big]\sin(\beta_4/2)\tfrac{\beta_2}{\beta_4},\ \  & \thg\rightarrow\pi ,
    \end{array}\right. \label{eq-3-36}\\
    \tan\varphi_A^\pm &\simeq \left\{\begin{array}{ll}
        \mp \tan(\beta_4/2) \big[1 - \big(\Re(R_0^\pm) \pm \Im(R_0^\pm) \tan(\beta_4/2)\big) \tfrac{g_2}{g_4} \big],\ \  & \thg\rightarrow 0 , \\
        \pm \tan(\beta_4/2) \big[1 + \big(\Re(R_0^\pm) \mp \Im(R_0^\pm) \tan(\beta_4/2)\big) \tfrac{g_2}{g_4} \big], & \thg\rightarrow\pi .
    \end{array}\right. \label{eq-3-37}
\end{align}
Note that $\Re(R_0^\pm) = \tR_0\cos(\tphi^\pm)$ corresponds to the part of $B^\pm$ parallel to the $A^\pm$ field and $\Im(R_0^\pm) = \tR_0\sin(\tphi^\pm)$ corresponds to the perpendicular component, i.e. the component of the electric field for $A$ which is parallel to the  magnetic field for $B$.

From the expressions above, we can infer the following:
\bi
\item
The radiation energy density is conserved when $R_A^\pm=1$, so the small term $\Delta R_A^\pm \equiv \big[\pm\Im(R_0^\pm) \cos(\beta_4/2) -  \Re(R_0^\pm)\sin(\beta_4/2)\big]\sin(\beta_4/2)\tfrac{\beta_2}{\beta_4}$ corresponds to the change in the amplitude of $A_\pm$ due to energy conversion. 
Note that we only keep terms linear in $\sin\thg$ here, whereas the spatially-averaged conversion fraction $P$ in Eq.~\eqref{eq:P_def} is $\cO(\sin^2\thg)$, which is subdominant in the coherent region.
The sign of $\Delta R_A^\pm$ determines the direction of energy transfer, i.e., whether energy is transferred from the $B$ field to the $A$ field or vice versa, and this direction depends on the effective initial phase $\tphi^\pm$. For example, if $\tphi^\pm=\pi/2$, $\Delta R_A^\pm$ must be positive when $|\beta_4(\eta)|<\pi$\footnote{Since $\beta_4(\eta)= 2g_4\int_{\eta_i}^{\eta}\phi'(\eta)d\eta$, the condition $|\beta_4(\eta)|<\pi$ must be satisfied at the beginning of the energy conversion.}, regardless of which gauge field is initially stronger or which coupling constant is larger.

\item If $\varphi_A^\pm = \mp\tfrac{1}{2}\beta_4\cdot\text{sgn}(\cos\thg)$, the cosmic birefringence angle becomes the same as the pure axion-photon coupling case ($\beta_A=\beta_1$), so $\Delta\beta_A\equiv \beta_A-\beta_1$ is determined by the $\cO(\beta_2/\beta_4)$ term, which corresponds to the small correction to this birefringence angle compared to the pure axion-photon case.
\ei

\begin{figure}
    \centering
    \includegraphics[width=0.45\linewidth]{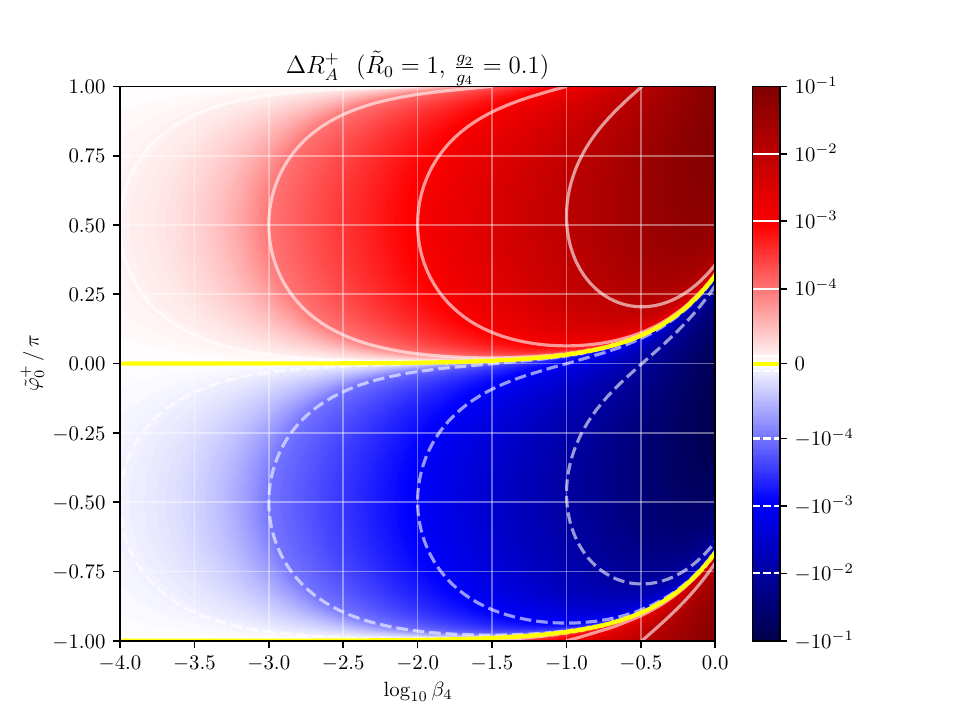}
    \includegraphics[width=0.45\linewidth]{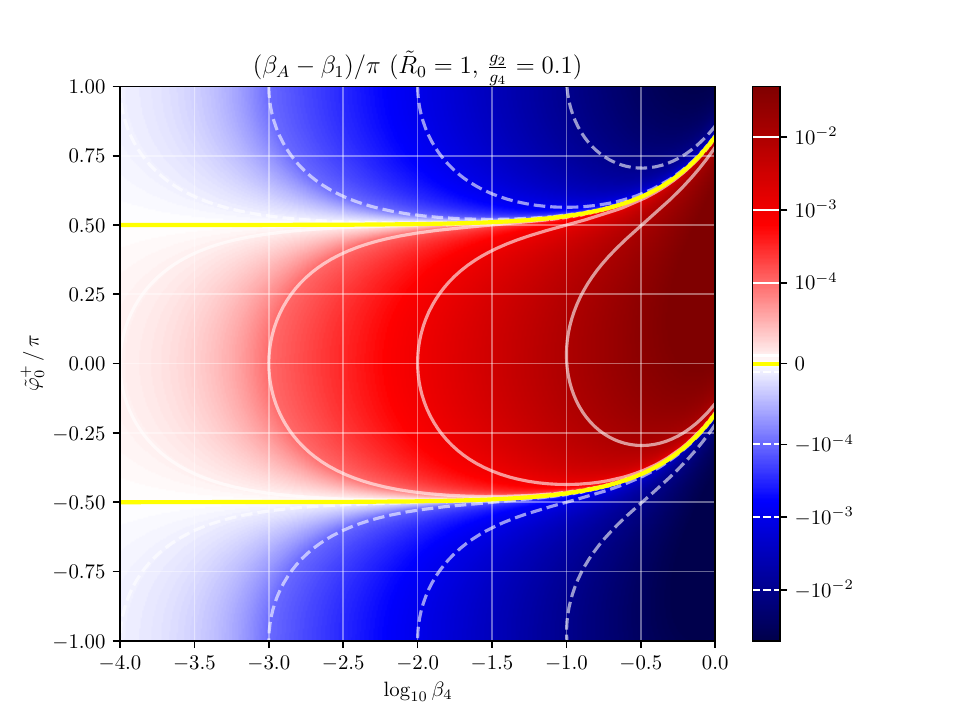}
    \caption{\tit{Left panel}: $\Delta R_A^+$ as a function of $\tdp$ and $\tphi^+$, for $\tR_0=1$ and $\frac{g_2}{g_4}=0.1$ and $g_3>g_1$, i.e. $\cos(\thg)>0$. The solid/dashed white curves correspond to the contour lines $\log_{10}|\Delta R_A^+|=-2,-3,-4,-5$ for positive/negative $\Delta R_A^+$, respectively, while the red curves indicate $\Delta R_A^+=0$. Note that $\Delta R_A^\pm\propto \tR_0\cdot\frac{g_2}{g_4}$, so this contour plot captures the full dependence of $\Delta R_A^+$ on these parameters. \tit{Right panel}: The difference between the photon cosmic birefringence angle and $\beta_1=2 g_1 I_\phi$, for the same parameter values as in the left panel. The yellow curves correspond to $\Delta\beta_A=0$.}
    \label{fig3}
\end{figure}

The value of $\Delta R_A^+$ and $\Delta\beta_A$ as a function of $\beta_4$ and $\tphi^+$ for $\tR_0=1$ and $g_2/g_4=0.1$ are shown in Fig.~\ref{fig3}. From Eqs.~\eqref{eq-3-36} and \eqref{eq-3-37}, we see that when $\beta_4$ is small, the perpendicular component of $B_\pm$, i.e., the imaginary part of $R_0$, makes up the dominant contribution to $\Delta R_A^\pm$; conversely,  the parallel component corresponding to $\Re(R_0)$ has the dominant impact on the correction to the cosmic birefringence angle, $\Delta\beta_A$.
This property is illustrated by the ``no correction" contour lines, i.e. the yellow curves in Fig.~\ref{fig3}. These curves are nearly horizontal at small $\beta_4$, but gradually bend upward with the increase of $\beta_4$.
In addition, at fixed $\beta_4$, the ``no correction'' contour in each panel corresponds to the maximal correction in the other panel. 
In particular, $\Delta R_A^+=0$ on the yellow curves in the left panel, i.e. $\tphi^\pm=\pm (\beta_4/2+n\pi)$, and $|\Delta\beta_A|$ reaches its maximum for a fixed $\beta_4$.
By contrast, when $\tan\tphi^\pm=\mp \cot(\beta_4/2)$ (yellow curves in the right panel), the correction in the birefringence angle is zero, while the energy conversion is maximized.

\subsubsection{Vanishing $g_1$} \label{ch-case5}

We also consider the case with negligible axion-photon coupling compared to the couplings with the dark photon. This is interesting because the axion-photon coupling is tightly constrained for a large range of the axion mass, but the interaction between different components in dark sector could still be strong.
In this case, $g_4=\sqrt{g_2^2+g_3^2}$, and we have from Eqs.~\eqref{eq-RA} and \eqref{eq:betaX_def}-\eqref{eq:betai_def} that
\begin{align}
    R_A^\pm(\eta)^2 &= 1 - (1-\tR_0^2)\Big(\frac{\beta_2}{\beta_4}\Big)^2 \sin^2\Big(\frac{\beta_4}{2}\Big) \\&\ \ \ + 2 \tR_0\Big[\pm  \sin(\tphi^\pm)\cos\Big(\frac{\beta_4}{2}\Big) - \cos(\tphi^\pm)\Big(\frac{\beta_3}{\beta_4}\Big)\sin\Big(\frac{\beta_4}{2}\Big) \Big] \Big(\frac{\beta_2}{\beta_4}\Big)\sin\Big(\frac{\beta_4}{2}\Big),
\end{align}
and
\begin{align}
    \beta_A &= \frac12\beta_3 - \arctan\Bigg[\frac{\beta_3 -\tR_0\cos(\tphi)\beta_2}{\tfrac{\beta_4}{\tan(\beta_4/2)} + \tR_0\sin(\tphi)\beta_2}\Bigg],
\end{align}
respectively. 
From this, we see that even if we do not have the standard coupling between the photon and axion, the interactions with the {\it dark} photon will lead to energy conversion and a nonzero cosmic birefringence angle. Also, even if we take $g_3=g_1=0$, we can still obtain non-zero $\Delta R_A^\pm$ and $\beta_A$ as long as $g_2$ is non-vanishing.

\section{Constraints from CMB observations}\label{sec:CMB}

It is interesting to consider constraints on the axion-photon-dark photon coupling from CMB spectral distortions and cosmic birefringence separately. However, the initial relative phase and angle are only well-defined when both the initial photon and dark photon fields are coherent. For the thermalized CMB photons, we use the spatially averaged quantities derived in subsection \ref{ch-case0}. 

\subsection{CMB spectral distortions}

Since we have established that axion-photon-dark photon interaction will lead to energy conversion between the photon and dark photon fields, we can constrain this model with spectral distortions. The energy conversion after recombination corresponds to y-distortion~\cite{Cyr:2024pme}; however, the photon–dark photon conversion produces a frequency-dependent mixture template instead of a Compton y spectral shape. Therefore, we need to compare the conversion template directly with monopole-spectrum data. Ref.~\cite{CMB-1996} shows that in the well-measured region, i.e. excluding the high-uncertainty tails, the CMB monopole spectrum deviates from a perfect blackbody spectrum by approximately $\Delta I/I_{bb} \lesssim 10^{-4}$, where $I_{bb}$ denotes the specific intensity of a perfect blackbody. We will adopt this constraint for the analytical estimate while using the full spectrum for Markov-Chain Monte Carlo (MCMC) analysis.

If we assume that the original spectra of $A$ and $B$ are thermalized, i.e. that their occupation numbers\footnote{We include the redshift effect in the temperature; $T_X=T_{X,i}/(z_{i}+1)$ corresponds to the temperature today.} obey the Bose-Einstein distribution with zero chemical potential
\begin{align}
    n_{T_{X}}(k) = \frac{1}{e^{k/{T_X}}-1},
\end{align}
where the $A$ field is the photon field corresponding to CMB photons. However, due to the energy conversion between photon and dark photon, the initial blackbody spectrum temperature $T_A$ could be slightly different from the temperature determined from the CMB spectral intensity \cite{CMB-1996}. Additionally, we assume that the temperature of the dark photon is $T_B=r \, T_A$, where the ratio $r$ should be smaller than $\approx0.5$\footnote{$\rho_{B}/\rho_{CMB}(t_{rec}) = (T_B/T_A)^4 = r^4 \leq 0.065\ \Rightarrow\ r<0.065^{1/4} \approx 0.5$} due to the constraint on the effective number of relativistic degrees of freedom $\Delta N_{\rm eff}$ from {\it Planck} CMB data \cite{Planck-2018}.
Then, we can derive the modulus of the ratio between the two initial fields
\begin{align}
    \tR_0 \equiv \left|\frac{B_0}{A_0}\right| &= \sqrt{\frac{u_{B}(k)}{u_A(k)}} = \sqrt{\frac{e^{k/T_{A}}-1}{e^{k/rT_{A}}-1}}\nn
    &\approx \left\{\begin{array}{ll}
        \sqrt{r} & k\ll T_{A} \\
        e^{\frac12(1-\frac1r)\frac{k}{T_A}} & k\gg T_{A} .
    \end{array}\right. \label{eq-R0-bb}
\end{align}
We can see that the ratio decreases monotonically with increasing $k$, approaching a constant value $\sqrt{r}$ as $k\rightarrow 0$ and decreasing exponentially toward zero as $k\rightarrow\infty$. Therefore, the ratio takes values between $0$ and $\sqrt{r}$.

The fractional spectral distortion after the axion-photon-dark photon interaction becomes\footnote{Note that the spatially averaged intensity is independent of the ``$\pm$'' helicity modes, so we ignore the ``$\pm$'' sign in the following expressions.}
\begin{align}
    \frac{\Delta I(k)}{I_{bb}(k)} &\,=\, \bigg| \frac{k^2|A_\pm(k)|^2}{k^2|A_{bb}^\pm(k)|^2}-1\bigg| \,\approx\, \left| (R_A^\pm)^2-1\right| 
    \,=\, \big| 1-\tR_0^2(k) \big| P,
\end{align}
where $I_{bb}=k^2|A_{bb}|^2$ is the best-fit blackbody intensity from today's observation, and we approximate $A_{bb}\approx A_0$ in the second line. Since $\Delta I/I\lesssim 10^{-4}$, this approximation is sufficiently accurate for our analytical expression. For the MCMC fitting, however, we use the exact expression for the spectral distortion.

Since the value of $\tR_0\in(0,\sqrt{r})$, the maximum $\Delta I/I$ should be lie at $k\rightarrow\infty$, i.e. for $\tR_0\rightarrow0$, and we can derive the analytical constraint
\begin{align}
    \frac{\Delta I}{I} \bigg|_{max} = P <10^{-4},
\end{align}
where $P\equiv \sin(\thg)^2 \sin(\beta_4/2)^2$ is the photon-dark photon energy conversion fraction in eq.~\eqref{eq-3-28}.
\begin{figure}
    \centering
    \includegraphics[width=0.46\linewidth]{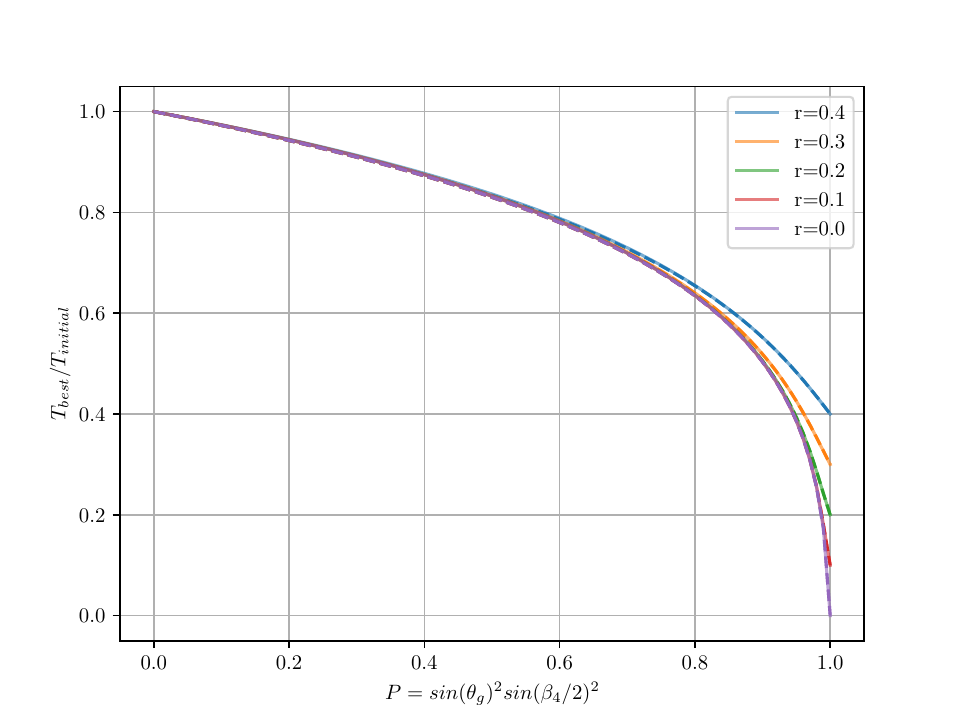}
    \includegraphics[width=0.46\linewidth]{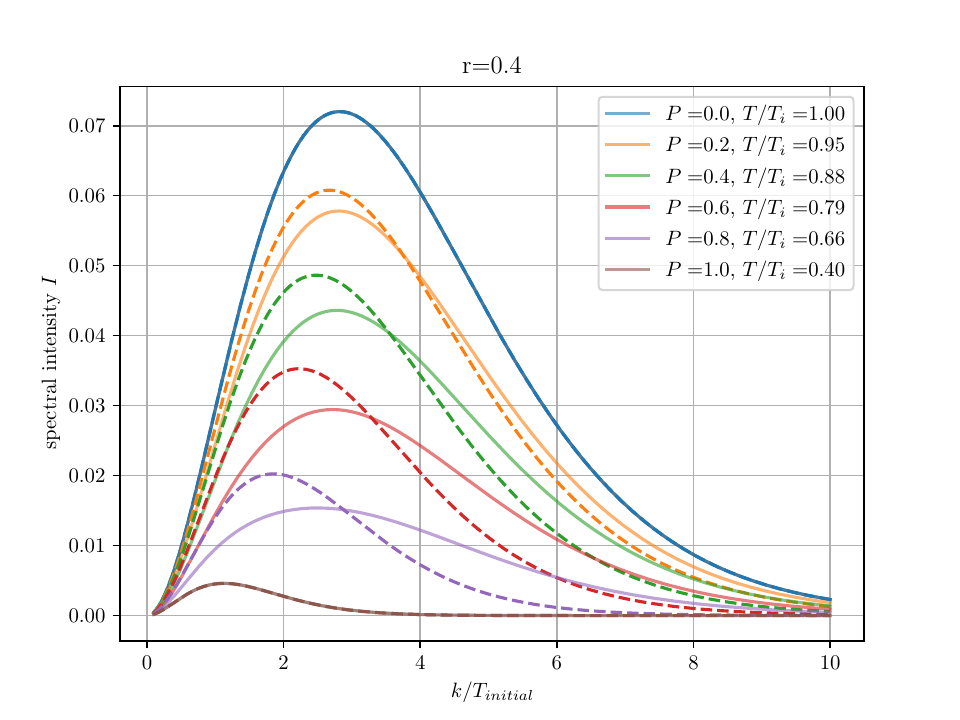}
    \caption{\tit{Top panel}: $T_{best}/T_{initial}$ as a function of $P=\sin(\thg)^2 \sin(\beta_4/2)^2$. The solid and dashed curves correspond to two different initial temperatures (but with the same $k$ bins), which does not affect the ratio. \tit{Bottom panel}: Comparing the spectrum intensity of CMB photons with the influence of dark photons (solid curves) with the best-fit blackbody spectrum (dashed curves). Note that the spectra with $P=0$ (blue curves) correspond to the original CMB spectrum, while $P=1$ (brown) curves correspond to the initial dark photon spectrum.}
    \label{fig3}
\end{figure}
On the other hand, by inserting the initial blackbody spectra, we obtain the exact expression for the spectral intensity
\begin{align}\label{eq:Iexact_def}
    I(k) &= \frac{k^3}{2\pi^2} \left(\frac{1-P}{e^{k/T_A}-1} + \frac{P}{e^{k/rT_A}-1}\right),
\end{align}
which depends on the three parameters $\{T_A,\ P,\ r\}.$ We can use MCMC to derive the posterior distributions of these parameters through Eq.~\eqref{eq:Iexact_def}. Note that $T_A=T_{A,i}(t_{\rm rec})/(z_{\rm rec}+1)$, the temperature of $A$ field today without the influence of the axion-photon-dark photon coupling, is different from $T_{\rm CMB}$. 
Fig.~\ref{fig3} shows how the energy conversion changes the spectral intensity and the corresponding temperature. 
The left panel shows the ratio of the observed temperature to the initial (redshifted) blackbody temperature, which significantly decrease with the increase of the energy conversion probability $P$, and when $P=1$, the temperature becomes the temperature of dark photons. This is consistent with the expression in Eq.~\eqref{eq-3-28} with $P=1$, i.e. $(\rho_A,\ \rho_B)=(\rho_{B,i},\ \rho_{A,i})$. The right panel illustrates the spectrum after energy conversion (solid curves) and the corresponding best-fit blackbody spectrum (dashed curves).

\begin{figure}
    \centering
    \includegraphics[width=0.7\linewidth]{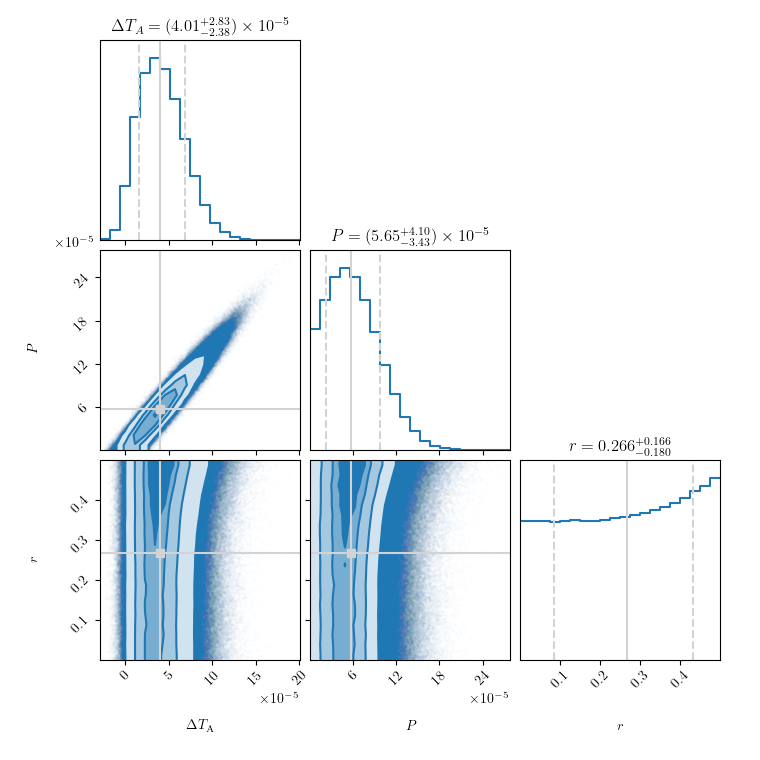}
    \caption{The posterior distribution of $\Delta T_A \equiv T_A-T_{CMB}$, $P$ and $r \equiv T_B/T_A$.}
    \label{fig11}
\end{figure}

We use MCMC to sample the posterior distribution of the parameters $\{T_A,\ P,\ r\}$, adopting uniform priors\footnote{Here we adopt the constraint on $r$ with $0\leq r\leq 0.5$ and $P\equiv \sin(\thg)^2\sin(\beta_4/2)^2>0$}. We use the CMB monopole spectrum measured by the FIRAS experiment over the frequency range $2.27-21.33\ {\rm cm}^{-1}$ \cite{CMB-1996}. The resulting constraints are shown in Fig.~\ref{fig11}.
We use the parameter $\Delta T_A \equiv T_A-T_{CMB}$, where $T_{CMB}=2.725K$ is the best-fit temperature of the observed  CMB spectrum. The posterior distributions of $\Delta T_A$ and $P$ have well-defined peaks, and the marginalized distributions indicate the best-fit values with the corresponding 68\% credible intervals as
\begin{align}
    \left\{\begin{array}{l}
        \Delta T_{A} = (4.01^{+2.83}_{-2.38})\times10^{-5} , \\[2mm]
        P = (5.65^{+4.10}_{-3.43})\times10^{-5} .
    \end{array}\right.
\end{align}
The posterior distribution of $r$ is relatively flat over the allowed parameter range $0\leq r\leq0.5$ with a mild preference for larger values of $r$. Thus, we do not show the best-fit value for it.
Moreover, the two-dimensional posterior distributions for $r$ show that this parameter is relatively insensitive to the values of $\Delta T_A$ and $P$.
It should be noted that we only consider the effect of the axion-photon-dark photon coupling and neglect other possible sources of spectral distortion in this MCMC analysis. Therefore, the posterior distribution of $P$ should be primarily interpreted as an upper bound on the photon-dark photon energy conversion fraction.

In particular, from Fig.~\ref{fig11}, values of $P>1.39\times10^{-4}$ are excluded at 95\% confidence level. On the other hand, $P=0$ remains allowed, and from the 2D posterior distribution, when $P=0$, i.e. the region close to the lower boundary of the first panel in the second row, the best-fit value of $T_A$ is very close to $T_{\rm CMB}$.

\subsection{Cosmic Birefringence}

The spatially averaged cosmic birefringence angle of the $A$ field is
\begin{align}\label{eq:birefangle_def}
    \langle\beta_A\rangle = \frac12(\beta_1+\beta_3) -  \arctan\big[\frac{\beta_3-\beta_1}{\beta_4}\tan\big(\beta_4/2\big)\big],
\end{align}
which we derive in Appendix~\ref{app-4}. This angle is independent of the initial conditions of the photon and dark photon fields, and is determined by the coupling constants and the axion field integral $I_\phi$. 

\begin{figure}
    \centering
    \includegraphics[width=0.45\linewidth]{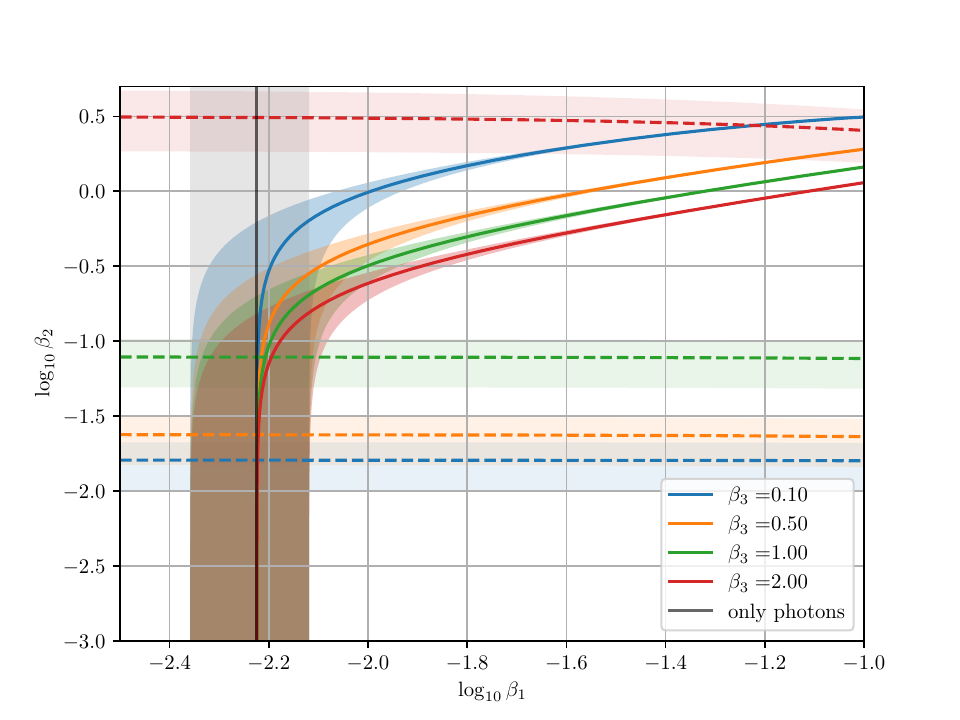}
    \caption{The regions in parameter space satisfying $\beta_A=0.342^\circ{}^{+0.094^\circ}_{-0.091^\circ} = (5.97^{+1.64}_{-1.59})\times10^{-3}\,[{\rm rad}]$ are shown in coloured bands. The coloured curves correspond to different values of $\beta_2$, while the gray curve represents the case with only the axion-photon interaction, i.e. $\beta_A=\beta_1$.}
    \label{fig12}
\end{figure}

In Fig.~\ref{fig12}, we show the constraints on these parameters obtained from the observed cosmic birefringence angle, $\beta_A=0.342^\circ{}^{+0.094^\circ}_{-0.091^\circ}$ (68\% C.L.) \cite{birefringence-2022}. The deeper shaded regions with solid curves correspond to the confidence regions of the CMB cosmic birefringence angle, and the lighter ones with dashed curves correspond to the constraints on the energy conversion fraction $P=(5.65^{+4.10}_{-3.43})\times 10^{-5}$ from spectral distortion. It should be noted that $\beta_A=0$ is excluded by 99.987\% credible level \cite{birefringence-2022}, while zero energy conversion between photon and dark photon is allowed. Thus, the allowed region in the parameter space should be the deeper shaded region below the corresponding dashed curves.
The four coloured curves correspond to the constraints with different values of $\beta_3$, while the black curve is the case with axion-photon coupling only, i.e. for $g_2=g_3=0$ and $\beta_A=\beta_1$. 

The constraint on the cosmic birefringence angle is significantly shifted with the increase of $\beta_2$. However, for small $\beta_3$, e.g. the $\beta_3=0.1$ case, the constraint on $\beta_2$ is strict, and the allowed region almost overlaps with the only axion-photon case. When $\beta_3$ is increased, the constraint on $\beta_2$ from spectral distortions is significantly relaxed and a wider range of $\beta_1$ is allowed. This means that the constraint on the axion-photon coupling constant $g_1$ can be relaxed in the presence of interactions with dark photons with larger values of $\beta_2$ and $\beta_3$. However, the allowed region becomes narrow for large $g_1$\footnote{In these regions, the parameter constraints are less sensitive to the uncertainty in the observed birefringence angle.}, indicating that these parameter values lie in a fine-tuned region, where the contributions from the individual interactions are relatively large but their combined effect on the birefringence angle $\beta_A$ remains small due to cancellations. In this sense, smaller values of $g_1$ and $g_3$ are less fine-tuned, as the observed birefringence angle does not require a delicate cancellation among the different contributions. 

In the case where the parameters $\beta_i$ are all small, which corresponds to the broad shaded regions in Fig.~\ref{fig12}, we can expand the expression for the birefringence angle \eqref{eq:birefangle_def} to third order as
\begin{align}
    \langle \beta_A\rangle &= \beta_1 - \frac{1}{24}(\beta_3-\beta_1)\beta_2^2 + \cO(\beta_i^5)\\
    &\simeq (1+\tfrac{P}{6})\beta_1 - \tfrac{P}{6}\beta_3,
\end{align}
where the energy conversion fraction $P\approx\beta_2^2/4$ for small $\beta_i$'s. Similarly, we can derive the global cosmic birefringence angle for dark photons, which, however, we cannot directly detect, as
\begin{align}
    \langle \beta_B\rangle &= (1+\tfrac{P}{6})\beta_3 - \tfrac{P}{6}\beta_1 + \cO(\beta_i^5),
\end{align}
where $\beta_3$ is the dark photon birefringence angle in the case when only the axion-dark photon interection $g_3\phi B\tilde{B}$ is non-zero. Therefore, in the limit $\beta_i\rightarrow 0$, we can call $-P/6$ the birefringence angle conversion fraction.

\section{Conclusions and Discussion} \label{ch5}

In this work, we studied the dynamics of photons and dark photons with the Chern-Simons type interactions $\phi F\tilde{F}$, $\phi B\tilde{F}$ and $\phi B\tilde{B}$, and their impact on CMB spectral distortions and cosmic birefringence. In a homogeneous axion background, these interactions change both the amplitudes and phases of the propagating gauge fields. The resulting energy conversion and polarization rotation, described by Eqs.~\eqref{eq-RA}--\eqref{eq-3-21}, allow the dark sector to affect two independently measured properties of the CMB. Keeping all three couplings is therefore useful for understanding how a constraint obtained from one observable changes when the other interactions are present. The result is sensitive to the initial polarization states of both the visible and dark sectors: for coherent fields, the relative phases affect both the conversion and the rotation, as illustrated in Fig.~\ref{fig1}; similarly, for initially incoherent radiation, the interference terms vanish after spatial averaging at the level of the radiation intensity, and the intensity transfer reduces to Eq.~\eqref{eq-3-28}. The averaged conversion is the same for both helicities, so it does not generate any net circular polarization when this is initially absent in both sectors.
Within the leading adiabatic approximation, the energy conversion fraction $P$ is independent of frequency; nevertheless, conversion between thermal populations at different temperatures generally produces a photon spectrum which deviates from a perfect blackbody. Equation~\eqref{eq:Iexact_def} shows that observed intensity contains contributions from both initial distributions, with each constituent carrying a weight based on the conversion fraction. Part of this change can be absorbed into a shift of the fitted temperature, making it necessary to vary $T_A$ together with $P$ and $r=T_B/T_A$. The remaining distortion then constrains the exchange of radiation between the two sectors.
The posterior distributions in Fig.~\ref{fig11} give $\Delta T_A=(4.01^{+2.83}_{-2.38})\times10^{-5}$ K and $P=(5.65^{+4.10}_{-3.43})\times10^{-5}$ at 68\% credibility, while the temperature ratio $r$ remains weakly constrained. The corresponding upper limit is $P<1.39\times10^{-4}$ at 95\% credibility. This upper limit is a useful result for discussing the allowed interactions.

For cosmic birefringence, Eq.~\eqref{eq:birefangle_def} shows that the measured rotation depends on all three couplings present in the action. In particular, the mixed interaction allows the axion-dark photon coupling to affect visible polarization, even though the dark radiation is not observed directly. The parameters constrained by this calculation are the combinations $\beta_i=2g_iI_\phi$, rather than the coupling constants separately. Since $I_\phi$ is the axion field difference between emission and observation, as defined in Eq.~\eqref{eq:Iphi_def}, translating these results into bounds on $g_i$ requires assumptions about the axion abundance, mass and phase.
The spectral constraint also limits how far the dark sector can change the usual interpretation of birefringence. When all the couplings $\beta_i$ are small, expanding Eq.~\eqref{eq:birefangle_def} gives $\langle\beta_A\rangle-\beta_1\simeq P(\beta_1-\beta_3)/6$, and the correction is therefore suppressed by the same conversion fraction constrained by the monopole spectrum. In this regime, the ordinary axion-photon result is a good approximation. Figure~\ref{fig12} illustrates how the birefringence constraint changes beyond this limit, but the spectral bound must be imposed on the same parameters, so regions with larger individual contributions become narrow when reproducing the observed angle.
A small value of $P$ does not, however, require every $\beta_i$ to be small: the oscillatory dependence in Eq.~\eqref{eq:P_def} produces the narrow minima visible in Fig.~\ref{fig2}, where the radiation can be close to its initial intensity even though $P$ may have been considerable at earlier times.

An interesting generalization of the results in this paper would be to consider the case of anisotropy, since spatial fluctuations of the axion and dark radiation would allow us to study direction-dependent rotation and intensity changes, together with their angular correlations. This could test whether a polarization signal and a spectral distortion are consistent with the same underlying interactions. More sensitive monopole measurements would further restrict the conversion fraction, while a joint analysis of spectral distortions and CMB polarization could determine how much freedom remains for a dark-sector contribution to cosmic birefringence.

\begin{acknowledgements}
HJ, DWJ, and NAN were supported by the Institute for Basic Science under the project code IBS-49 R018-D3. NAN also acknowledges
support from PSL/Observatoire de Paris.
\end{acknowledgements}

\appendix

\section{Kinetic-diagonalization of the gauge fields} \label{app-1}

The origin action with the kinetic mixing term is
\begin{equation}\label{eq:lagr_main}
\begin{aligned}
    S=\int d^4x\sqrt{-g}\Bigg[&\frac{R}{2\kappa} -\frac{1}{4} \hat{B}_{\mu \nu } \hat{B}^{\mu \nu } -  \frac{1}{4} \hat{F}_{\mu \nu } \hat{F}^{\mu \nu } + \alpha \hat{B}_{\mu \nu } \hat{F}^{\mu \nu } + \hat{g}_{1}\phi \hat{F}_{\mu \nu } \tilde{\hat{F}}^{\mu \nu } \\& + \hat{g}_{2}\phi \hat{B}_{\mu \nu } \tilde{\hat{F}}^{\mu \nu }+ \hat{g}_{3}\phi \hat{B}_{\mu \nu } \tilde{\hat{B}}^{\mu \nu } -  V(\phi) - \frac12 \partial_{\mu }\phi \partial^{\mu }\phi \Bigg],
\end{aligned}
\end{equation}
where $\kappa\equiv8\pi G$, $R$ is the Ricci scalar. The coefficient $\alpha$ is dimensionless $[\alpha]=0$, and the $\hat{g}_i$'s have dimension of inverse mass, $[\hat{g}_i]=-1$.
By an invertible linear field redefinition and canonical normalization of the kinetic matrix, the action \eqref{eq:lagr_main} can be rewritten by the following field redefinition and canonical renormalization
\begin{equation}
    \hat{B}_{\mu\nu} \to B_{\mu\nu}+\frac{2\alpha}{\sqrt{1-4\alpha^2}} F_{\mu\nu}, \quad     \hat{F}_{\mu\nu}\to \frac{1}{\sqrt{1-4\alpha^2}}F_{\mu\nu},
\end{equation}
which is equivalent to
\begin{equation}
    \begin{pmatrix}
        B_{\mu\nu}\\ F_{\mu\nu}
    \end{pmatrix}
    =
    \begin{pmatrix}
        1 & -2\alpha \\
        0 & \sqrt{1-4\alpha^2}
            \end{pmatrix}
        \begin{pmatrix}
        \hat{B}_{\mu\nu}\\ \hat{F}_{\mu\nu},
        \end{pmatrix}
\end{equation}
where we note that $|\alpha|<1/2$ must be satisfied. Furthermore, we define the constants
\begin{equation}
    \begin{aligned}
g_1 =& \frac{[\hat{g}_1+2\alpha(\hat{g}_2+2\alpha\hat{g}_3)]}{1-4\alpha^2}, \\
g_2 =& \frac{\hat{g}_2+4\alpha\hat{g}_3}{\sqrt{1-4\alpha^2}}, \\
g_3 =& \hat{g}_3,
    \end{aligned}
\end{equation}
after which we get the action in eq.~\eqref{eq-L}.

\section{the expression of the eigenvectors}\label{app-2}

For the real symmetric matrix $C$, we can diagonalize it with an orthogonal matrix $M$ 
\begin{align}
    C = MDM^{-1},
\end{align}
where $D=diag(\lambda_1,\lambda_2)$, and the corresponding eigenvectors are
\begin{align}
v_b = \left(\begin{array}{c}
    -2g_2 \\
    4g_1 - \lambda_b
\end{array}\right)
= \left(\begin{array}{c}
    -2g_2 \\
    2g_1-2g_3 \mp 2g_4
\end{array}\right)
\end{align}

And to get $M$, we need to normalize the eigenvector $u_b = v_b/||v_b||$, where $||v_b||^2 = 8\small\sqrt{(g_1-g_3)^2 +g_2^2}\big[ \small\sqrt{(g_1-g_3)^2 +g_2^2} \mp (g_1-g_3)  \big]$ (where $+/-$ signs correspond to $b=1$ and 2).

\begin{align}
\lambda_1 &= 2\big( g_1 + g_3 + \small\sqrt{(g_1-g_3)^2 + g_2^2} \big) \\
\lambda_2 &= 2\big( g_1 + g_3 - \small\sqrt{(g_1-g_3)^2 + g_2^2} \big),
\end{align}

\begin{align}
    &\vec{A}_\pm = M \vec{\mcal{A}}_\pm\\
    \Rightarrow\ \ 
    &a_1 = -\frac{2g_2}{||v_1||},\ \ \ a_2 = -\frac{2g_2}{||v_2||}\\
    &b_1 = \frac{4g_1-\lambda_1}{||v_1||},\ b_2 = \frac{4g_1-\lambda_2}{||v_2||}
\end{align}

According the definition of the parameter set $\{ g_0,\ g_4,\ \theta_g\}$ before, we have
\begin{align}
    g_1 &= (g_0 - g_4\cos\theta_g)/2\\
    g_2 &= g_4\sin\theta_g \\
    g_3 &= (g_0 + g_4\cos\theta_g)/2 \\
    \lambda_{1,2} &= 2(g_0\pm g_4)\\
    ||v_{1,2}||^2 &= 8g_4^2(1\pm\cos\thg) 
\end{align}
where $\lambda_1$ and $\lambda_2$ correspond to the ``$+$" and ``$-$" sign separately.
Then, the normalized eigenvectors can be calculated as
\begin{align}
    ||v_1||^2 &= 8g_4^2(1+\cos\theta_g) = 16g_4^2\cos^2(\theta_g/2)\\
    \Rightarrow\ \ \ \ \cos\alpha_{1} &\equiv a_{1} = -\frac{2g_2}{||v_{1}||} = -\frac{\sin\theta_g}{2\cos(\theta_g/2)} = -\sin(\theta_g/2)\\
    \sin\alpha_{1} &\equiv b_{1} = \frac{4g_1-\lambda_1}{||v_{1}||} = -\frac{\cos\theta_g+1}{2\cos(\theta_g/2)} = -\cos(\theta_g/2)
\end{align}
and
\begin{align}
    ||v_2||^2 &= 8g_4^2(1-\cos\theta_g) = 16g_4^2\sin^2(\theta_g/2)\\
    \Rightarrow\ \ \ \ \cos\alpha_{2} &\equiv a_{2} = -\frac{2g_2}{||v_{2}||} = -\frac{\sin\theta_g}{2\sin(\theta_g/2)} = -\cos(\theta_g/2)\\
    \sin\alpha_{2} &\equiv b_{2} = \frac{4g_1-\lambda_2}{||v_{2}||} = \frac{1-\cos\theta_g}{2\sin(\theta_g/2)} = \sin(\theta_g/2)
\end{align}
Note that as $g_2>0$, $\theta_g\in[0,\pi]$, so $\cos/\sin(\theta_g/2)\geq0$.

Then, we can derive the relation between $\alpha_1,\ \alpha_2$ and $\theta_g$:
\begin{align}
    &\alpha_1 = \frac{3}{2}\pi - \frac{\theta_g}{2}\ ,\ \ 
    \alpha_2 = \pi - \frac{\theta_g}{2}\\
    \Rightarrow\ \ \ \ 
    &2\alpha_1 = \pi-\theta_g\ ,\ \ 
    2\alpha_2 = -\theta_g
\end{align}
Note that here we modulo $2\pi$ in the calculation.

So the orthogonal matrix can be expressed as
\begin{align}
    M = 
    \left(\begin{array}{cc}
        -\sin(\thg/2) & -\cos(\thg/2) \\
        -\cos(\thg/2) & \sin(\thg/2)
    \end{array}\right)
\end{align}

\section{Compute $A^\mu$ and $B^\mu$} \label{app-3}

Let us first discuss the phase of $A_\pm$ and $B_\pm$:

For simplification, we consider linearly polarized electromagnetic waves which propagate along the $z$-axis, and the helicity basis is defined as $\hat{\mbf{\epsilon}}_\pm = \frac{1}{\sqrt{2}}(\hat{\mbf{x}}\pm i\hat{\mbf{y}})$. Then, we have
\begin{align}
    A_\pm = \tfrac{1}{\sqrt{2}}(A_x\mp iA_y) = \tfrac{1}{\sqrt{2}}|\mbf{A}|e^{\pm  i{\tphi}{}_{,A} + i{\tPhi}{}_{,A}}\\
    B_\pm = \tfrac{1}{\sqrt{2}}(B_x\mp iB_y) = \tfrac{1}{\sqrt{2}}|\mbf{B}|e^{\pm i{\tphi}{}_{,B} + i{\tPhi}{}_{,B}}
\end{align}
where $\tphi{}_{,A}=-\arctan\frac{A_y}{A_x}$ and $\tphi{}_{,B}=-\arctan\frac{B_y}{B_x}$ (the polar angle). Note that here we assume that at recombination, the relative phase between $A_0^+$ and $A_0^-$ is zero.

Then, it is reasonable to assume that 
\begin{align}
    B_0^{\pm}/A_0^{\pm} &\equiv R_0^{\pm} =  \tilde{R}_0 e^{i\tphi^\pm},\\
    \Rightarrow\ \ A_0^{\pm}/B_0^{\pm} &\,\,\equiv (R_0^{\pm})^{-1} =  \tilde{R}_0^{-1} e^{-i\tphi^\pm}
\end{align}
where $\tilde{R}_0 = |\mbf{B}(t_{rec})|/|\mbf{A}(t_{rec})|$ is the modulus of $R_0$ and $\tphi^\pm = \pm(\tphi{}_{,B}-\tphi{}_{,A})  + (\tPhi{}_{,B}-\tPhi{}_{,A}) = \pm \tphi +\tPhi$ is the relative phase (corresponds to the relative angle between the two linearly polarized wave).

Then, we have
\begin{align}
    &\vec{A}_\pm = M \vec{\mcal{A}}_\pm,
\end{align}
where
\begin{align}
    M = 
    \left(\begin{array}{cc}
        \cos\alpha_1 & \cos\alpha_2 \\
        \sin\alpha_1 & \sin\alpha_2
    \end{array}\right)
    = \left(\begin{array}{cc}
        -\sin(\thg/2) & -\cos(\thg/2) \\
        -\cos(\thg/2) & \sin(\thg/2)
    \end{array}\right).
\end{align}
\begin{align}
    A_\pm &= \cos\alpha_1 \mcal{A}_1^\pm + \cos\alpha_2 \mcal{A}_2^\pm \nn
          &= \cos\alpha_1 \mcal{A}_{0,1}^\pm e^{-i\int\omega_1^\pm(\eta) \,d\eta} + \cos\alpha_2 \mcal{A}_{0,2}^\pm e^{-i\int\omega_2^\pm(\eta) \,d\eta} \\
    B_\pm &= \sin\alpha_1 \mcal{A}_1^\pm + \sin\alpha_2 \mcal{A}_2^\pm \nn
          &= \sin\alpha_1 \mcal{A}_{0,1}^\pm e^{-i\int\omega_1^\pm(\eta) \,d\eta} + \sin\alpha_2 \mcal{A}_{0,2}^\pm e^{-i\int\omega_2^\pm(\eta) \,d\eta} .
\end{align}

Inserting the expression of $\mcal{A}_b$ as a function of $\alpha_1$ and $\alpha_2$, i.e. $\mcal{A}_b^\pm = \cos\alpha_b A_\pm + \sin\alpha_b B_\pm$
\begin{align}
    A_\pm &= \cos\alpha_1 (\cos\alpha_1 A_0^\pm + \sin\alpha_1 B_0^\pm) e^{-i\int\omega_1^\pm(\eta) \,d\eta} + \cos\alpha_2 (\cos\alpha_2 A_0^\pm + \sin\alpha_2 B_0^\pm) e^{-i\int\omega_2^\pm(\eta) \,d\eta} \\
    B_\pm &= \sin\alpha_1 (\cos\alpha_1 A_0^\pm + \sin\alpha_1 B_0^\pm) e^{-i\int\omega_1^\pm(\eta) \,d\eta} + \sin\alpha_2 (\cos\alpha_2 A_0^\pm + \sin\alpha_2 B_0^\pm) e^{-i\int\omega_2^\pm(\eta) \,d\eta}.
\end{align}
\begin{align}
\Rightarrow\ \     A_\pm 
    &= \big(\frac12+\frac12\cos2\alpha_1 \big)A_0^\pm e^{-i\int\omega_1^\pm(\eta) \,d\eta} + \frac12\sin(2\alpha_1)A_0^\pm \tR_0 e^{i\tphi^\pm} e^{-i\int\omega_1^\pm(\eta) \,d\eta} \nonumber\\
    &\ \ \ \ + \big(\frac12+\frac12\cos2\alpha_2 \big)A_0^\pm e^{-i\int\omega_2^\pm(\eta) \,d\eta} + \frac12\sin(2\alpha_2)A_0^\pm \tR_0 e^{i\tphi^\pm} e^{-i\int\omega_2^\pm(\eta) \,d\eta} \\
    B_\pm 
    &= \big(\frac12-\frac12\cos2\alpha_1 \big)B_0^\pm e^{-i\int\omega_1^\pm(\eta) \,d\eta} + \frac12\sin(2\alpha_1)B_0^\pm \tR_0^{-1}e^{-i\tphi^\pm} e^{-i\int\omega_1^\pm(\eta) \,d\eta} \nonumber\\
    &\ \ \ \ + \big(\frac12-\frac12\cos2\alpha_2 \big)B_0^\pm e^{-i\int\omega_2^\pm(\eta) \,d\eta} + \frac12\sin(2\alpha_2)B_0^\pm \tR_0^{-1}e^{-i\tphi^\pm} e^{-i\int\omega_2^\pm(\eta) \,d\eta} .
\end{align}

In the following, we set
\begin{align}
    \tom_1 &\equiv - \int\omega_1^\pm(\eta) \,d\eta\\
    \tom_2 &\equiv -\int\omega_2^\pm(\eta) \,d\eta.
\end{align}
Note that there is a ``$-$" sign before the integral corresponds to the ``$-$" sign in the time-dependent phase.

\begin{align}
    \Rightarrow\ \ 
    \Re(A_\pm) 
    &= A_0 \Big[ \Big(\tfrac12+\tfrac12\cos2\alpha_1+\tfrac{\tR_0}{2}\sin2\alpha_1\cos\tphi^\pm\Big) \cos(\tom_1) - \tfrac{\tR_0}{2}\sin2\alpha_1 \sin\tphi^\pm \sin(\tom_1) \nonumber\\
    &\ \ \ \ \ \ \ \ \ + \Big(\tfrac12+\tfrac12\cos2\alpha_2+\tfrac{\tR_0}{2}\sin2\alpha_2 \cos\tphi^\pm \Big) \cos(\tom_2) - \tfrac{\tR_0}{2} \sin2\alpha_2 \sin\tphi^\pm \sin(\tom_2) \Big] \\
    \Im(A_\pm) 
    &= A_0 \Big[ \Big(\tfrac12+\tfrac12\cos2\alpha_1+\tfrac{\tR_0}{2} \sin2\alpha_1 \cos\tphi^\pm \Big) \sin(\tom_1) + \tfrac{\tR_0}{2} \sin2\alpha_1 \sin\tphi^\pm \cos(\tom_1) \nonumber\\
    &\ \ \ \ \ \ \ \ \ + \Big(\tfrac12+\tfrac12\cos2\alpha_2+\tfrac{\tR_0}{2} \sin2\alpha_2 \cos\tphi^\pm \Big) \sin(\tom_2) + \tfrac{\tR_0}{2} \sin2\alpha_2 \sin\tphi^\pm \cos(\tom_2) \Big]\\
    \Re(B_\pm) 
    &= B_0 \Big[ \Big(\tfrac12-\tfrac12\cos2\alpha_1+\tfrac{\tR_0^{-1}}{2}\sin2\alpha_1\cos\tphi^\pm\Big) \cos(\tom_1) + \tfrac{\tR_0^{-1}}{2}\sin2\alpha_1 \sin\tphi^\pm \sin(\tom_1) \nonumber\\
    &\ \ \ \ \ \ \ \ \ + \Big(\tfrac12-\tfrac12\cos2\alpha_2+\tfrac{\tR_0^{-1}}{2}\sin2\alpha_2 \cos\tphi^\pm \Big) \cos(\tom_2) + \tfrac{\tR_0^{-1}}{2} \sin2\alpha_2 \sin\tphi^\pm \sin(\tom_2) \Big] \\
    \Im(B_\pm) 
    &= B_0 \Big[ \Big(\tfrac12-\tfrac12\cos2\alpha_1+\tfrac{\tR_0^{-1}}{2}\sin2\alpha_1\cos\tphi^\pm\Big) \sin(\tom_1) - \tfrac{\tR_0^{-1}}{2}\sin2\alpha_1 \sin\tphi^\pm \cos(\tom_1) \nonumber\\
    &\ \ \ \ \ \ \ \ \ + \Big(\tfrac12-\tfrac12\cos2\alpha_2+\tfrac{\tR_0^{-1}}{2}\sin2\alpha_2 \cos\tphi^\pm \Big) \sin(\tom_2) - \tfrac{\tR_0^{-1}}{2} \sin2\alpha_2 \sin\tphi^\pm \cos(\tom_2) \Big].
\end{align}


Let us try to rewrite the above expression. Setting the average frequency and the half frequency difference related to the two eigenfrequencies $\omega_1$ and $\omega_2$ as 
\begin{align}
    \omega^\pm(\eta) &\equiv \frac12 (\omega_1^\pm + \omega_2^\pm) = k_c\pm \frac14 (\lambda_1+\lambda_2)\phi' = k_c \pm  g_0\phi' \\
    \delta^\pm(\eta) &\equiv \frac12(\omega_1-\omega_2) = \pm \frac14 (\lambda_1-\lambda_2)\phi' = \pm  g_4\phi',
\end{align}
and 
\begin{align}
    \tilde\omega^\pm(\eta) &\equiv -\int\omega^\pm(\eta) \,d\eta = -k_c\cdot\eta(\eta) \mp g_0I_\phi(\eta) \\
    \tilde\delta^\pm(\eta) &\equiv -\int\delta^\pm(\eta) \,d\eta = \mp g_4 I_\phi(\eta).
\end{align}

Then, we have
\begin{align}
    \cos(\tom_1) &= \cos(\tom)\cos\big(\td\big) -\sin(\tom)\sin\big(\td\big)\\
    \sin(\tom_1) &= \sin(\tom)\cos\big(\td\big) +\cos(\tom)\sin\big(\td\big)\\
    \cos(\tom_2) &= \cos(\tom)\cos\big(\td\big) +\sin(\tom)\sin\big(\td\big)\\
    \sin(\tom_2) &= \sin(\tom)\cos\big(\td\big) -\cos(\tom)\sin\big(\td\big).
\end{align}
The real and imaginary parts of $A_\pm$ and $B_\pm$ become
\begin{align}
    Re(A_\pm)/A_0^{\pm}
    &= \Big[ \cos\big(\td\big) - \tR_0\sin\tphi^\pm\sin(\theta_g)\sin\big(\td\big) \Big] \cos(\tom) + \Big[\Big(\cos\theta_g - \tR_0\cos\tphi^\pm\sin\theta_g\Big)\sin\big(\td\big) \Big] \sin(\tom) \nn
    &= R_A^\pm(\eta) \cos(\tom - \varphi_A^\pm(\eta)) \\
        Im(A_\pm)/A_0^{\pm} 
    &= \Big[\cos\big(\td\big) - \tR_0\sin\tphi^\pm\sin(\theta_g)\sin\big(\td\big)\Big] \sin(\tom) - \Big[ \Big(\cos(\theta_g) - \tR_0\cos\tphi^\pm\sin(\theta_g)\Big)\sin\big(\td\big) \Big] \cos(\tom) \nn
    &= R_A^\pm(\eta) \sin(\tom - \varphi_A^\pm(\eta)) \\    
    Re(B_\pm)/B_0^{\pm} 
    &= \Big[ \cos\big(\td\big) + \tR_0^{-1}\sin\tphi^\pm\sin(\theta_g)\sin\big(\td\big) \Big] \cos(\tom) - \Big[\Big(\cos\theta_g + \tR_0^{-1}\cos\tphi^\pm\sin\theta_g\Big)\sin\big(\td\big) \Big] \sin(\tom) \nn
    &= R_B^\pm(\eta) \cos(\tom - \varphi_B^\pm(\eta)) \\
    Im(B_\pm)/B_0^{\pm} 
    &= \Big[\cos\big(\td\big) + \tR_0^{-1}\sin\tphi^\pm\sin(\theta_g)\sin\big(\td\big)\Big] \sin(\tom) + \Big[ \Big(\cos(\theta_g) + \tR_0^{-1}\cos\tphi^\pm\sin(\theta_g)\Big)\sin\big(\td\big) \Big] \cos(\tom) \nn
    &= R_B^\pm(\eta) \sin(\tom - \varphi_B^\pm(\eta)), 
\end{align}
where
\begin{align}
    R_A^\pm(\eta) \cos(\varphi_A^\pm(\eta)) \,&=\, \cos\big(\td\big) - \tR_0\sin\tphi^\pm\sin(\theta_g)\sin\big(\td\big) \\
    R_A^\pm(\eta) \sin(\varphi_A^\pm(\eta)) \,&=\,  \Big(\cos(\theta_g) - \tR_0\cos\tphi^\pm\sin(\theta_g)\Big)\sin\big(\td\big)\\
    R_B^\pm(\eta) \cos(\varphi_B^\pm(\eta)) \,&=\, \cos\big(\td\big) + \tR_0^{-1}\sin\tphi^\pm\sin(\theta_g)\sin\big(\td\big)\\
    R_B^\pm(\eta) \sin(\varphi_B^\pm(\eta)) \,&=\, -\Big(\cos(\theta_g) + \tR_0^{-1}\cos\tphi^\pm\sin(\theta_g)\Big)\sin\big(\td\big).
\end{align}

Thus, we have 
\begin{align}
    A_\pm &= A_0^\pm R_A^\pm(\eta) \exp\big[i\big(\tom - \varphi_A^\pm(\eta)\big)\big]\\
    B_\pm &= B_0^\pm R_B^\pm(\eta) \exp\big[i\big(\tom - \varphi_B^\pm(\eta)\big)\big],
\end{align}
which are slowly varying wave functions.

Inserting the expression of $\tom(\eta)$ and $\td(\eta)$, we get the result in Sec.~\ref{ch3}.


Note that the total energy of photons and dark photons remains constant
\begin{align}
    E_{tot}(k_c) &= k_c^2|A_\pm|^2 + k_c^2|B_\pm|^2 = k_c^2\Big(|A_0^\pm|^2R_A^2 + |B_0^\pm|^2R_B^2 \Big) \nn
    &= k_c^2 (|A_0^\pm|^2+|B_0^\pm|^2)\nn
    &\ \ \ \ \cdot \big[\cos\big(\td\big)^2 + \cos(\thg)^2\sin\big(\td\big)^2 + \sin\big(\tphi^\pm)^2\sin(\thg)^2\sin\big(\td\big)^2 + \cos\big(\tphi^\pm\big)^2\sin(\thg)^2\sin\big(\td\big)^2\big]\nn
    &= k_c^2|A_0^\pm|^2 + k_c^2|B_0^\pm|^2.
\end{align}
This is computed at leading adiabatic order, which is valid for $k_c\gg m$ where the WKB approximation is satisfied.

\section{Average value of the birefringence angle} \label{app-4}

With the linear polarization Stokes parameter $P_A = Q_A+iU_A = 2 A_+ A_-^*$, the observed birefringence angle is
\begin{align}
    \beta_A^{obs} &= -\frac{1}{2}\arg\Big(\frac{P_A}{P_{A_0}}\Big) = -\frac{1}{2} \arg\Big(\frac{A_+ A_-^*}{A_0^+ A_0^{-*}}\Big) \\
    &= -\frac{1}{2} \arg\Big(R_A^+ R_A^- \exp\big[-i(2 g_0 I_\phi+\varphi_A^+-\varphi_A^-)\big]\Big) \\
    &= g_0 I_\phi + \frac{1}{2}(\varphi_A^+-\varphi_A^-) = \beta_A
\end{align}
Thus, the spatially-averaged birefringence angle is
\begin{align}
    \langle\beta_A \rangle &= g_0 I_\phi + \frac{1}{2}\langle\varphi_A^+\rangle - \frac{1}{2}\langle\varphi_A^-\rangle
\end{align}

To compute the spatially averaged $\varphi_A^\pm$, we need to compute the integral
\begin{align}
    I_1 = \frac{1}{2\pi}\int_0^{2\pi} d\psi\, \arctan\Big(\frac{a_1+a_2\cos\psi}{b_1+b_2\sin\psi}\Big),
\end{align}
where $a_2=b_2\equiv c_2$.
This integrand corresponds to the angular coordinates of points on a circle centred at $(a_1, b_1)$ with radius $c_2$.

Let us define
\begin{align}
    F(a_1, b_1, c_2) = \int_0^{2\pi} d\psi\, \arctan\Big(\frac{a_1+a_2\cos\psi}{b_1+b_2\sin\psi}\Big),
\end{align}
Then, we have
\begin{align}
    \frac{\partial F}{\partial a_1} &= \int_0^{2\pi} d\psi\frac{b_1+b_2\sin\psi}{(a_1+a_2\cos\psi)^2+(b_1+b_2\sin\psi)^2}\\
    &= \int_0^{2\pi} d\psi \frac{b_1+c_2\sin\psi}{a_1^2+b_1^2+c_2^2 + 2c_2(a_1\cos\psi + b_1\sin\psi)}.
\end{align}
Define $c_1=\sqrt{a_1^2+b_1^2}$, and $a_1=c_1\cos\sigma,\ b_1=c_1\sin\sigma$. Then, the function becomes
\begin{align}
    \frac{\partial F}{\partial a_1} &= \int_0^{2\pi} d\psi\frac{b_1+c_2\sin\psi}{c_1^2+c_2^2+2 c_1 c_2 \cos(\psi-\sigma)}\\
    &= \sin\sigma \int_{-\sigma}^{2\pi-\sigma}dx \frac{c_1+c_2\cos x}{c_1^2+c_2^2+2 c_1 c_2 \cos x} + \cos\sigma \int_{-\sigma}^{2\pi-\sigma}dx \frac{c_2\sin x}{c_1^2+c_2^2+2 c_1 c_2 \cos x}.
\end{align}
The second term vanishes because the integrand is an odd function,
\begin{equation}
    T_2 = \cos\sigma \int_{-\pi}^{\pi} dx\frac{c_2\sin x}{c_1^2+c_2^2+2 c_1 c_2 \cos x} =0.
\end{equation}
Then, we have
\begin{align}
    \frac{\partial F}{\partial a_1} &= \sin\sigma \int_0^{2\pi} dx\frac{c_1+c_2\cos x}{c_1^2+c_2^2+2 c_1 c_2 \cos x} \\
    &= \sin\sigma \int_0^{2\pi} dx \Big[ \frac{1}{2c_1} + \frac{c_1^2-c_2^2}{2c_1} \frac{1}{c_1^2+c_2^2+2 c_1 c_2 \cos x} \Big]\\
    &= \sin\sigma\Big[\frac{\pi}{c_1} + \frac{c_1^2-c_2^2}{2c_1}\frac{2\pi}{|c_1^2-c_2^2|} \Big] = \frac{\pi b_1}{c_1^2}[1+\text{sgn}(c_1^2-c_2^2)]\\
    &= \left\{\begin{array}{ll}
        2\pi b_1/c_1^2, & c_1>|c_2| \\
        0, & c_1<|c_2|
    \end{array}\right..
\end{align}
For $a_1=0$, we have
\begin{align}
    F(a_1=0,b_1,c_2) &= \int_0^{2\pi} d\psi\, \arctan\Big(\frac{a_2\cos\psi}{b_1+b_2\sin\psi}\Big) = 0. 
\end{align}

We express these parameters in terms of the original model parameters as follows:
\begin{align}
    a_1 &= \cos(\thg)\sin\big(\td\big) \\
    b_1 &= \cos\big(\td\big) \\
    \Rightarrow\ \ c_1^2 &= 1- \sin(\thg)^2\sin\big(\td\big)^2\\
    c_2 &= -\tR_0 \sin(\thg)\sin\big(\td\big),
\end{align}
and with the small value of $s=\sin(\thg)\sin\big(\tdp\big) < 10^{-2}$ and for moderate values of $\tR_0$, we have $c_2^2< b_1^2\leq c_1^2$, so 
\begin{align}
    I = \frac{1}{\cancel{2\pi}} \int_0^{a_1} da_1' \frac{\cancel{2\pi}\,b_1}{a_1'^2+b_1^2} = \arctan\Big(\frac{a_1}{b_1}\Big) = \arctan\Big[\cos(\thg)\tan\big(\td\big)\Big].
\end{align}

For the averaged phase $\langle\varphi_A^\pm\rangle$, we have
\begin{align}
    \langle\varphi_A^\pm\rangle &= \frac{1}{(2\pi)^2}\int_0^{2\pi}d\tphi \int_0^{2\pi}d\tPhi \arctan\Big[\frac{\cos(\thg)\sin\big(\td\big)-\tR_0\cos\big(\tPhi\pm\tphi\big) \sin(\thg)\sin\big(\td\big)}{\cos\big(\td\big)-\tR_0 \sin\big(\tPhi\pm\tphi\big) \sin(\thg)\sin\big(\td\big)}\Big] \\
    &= \frac{1}{(2\pi)^2}\int_0^{2\pi}d\tphi \int_{\pm\tphi}^{2\pi\pm\tphi}d\tphi^\pm \arctan\Big[\frac{\cos(\thg)\sin\big(\td\big)-\tR_0\cos(\tphi^\pm) \sin(\thg)\sin\big(\td\big)}{\cos\big(\td\big)-\tR_0 \sin(\tphi^\pm) \sin(\thg)\sin\big(\td\big)}\Big]\\
    &= \frac{1}{2\pi}\int_0^{2\pi}d\tphi \arctan\Big[\cos(\thg)\tan\big(\td\big)\Big]
    = \arctan\Big[\cos(\thg)\tan\big(\td\big)\Big]\\
    &= \mp \arctan\Big[\cos(\thg)\tan\big(g_4 I_\phi(\eta)\big)\Big].
\end{align}

Thus, the birefringence angle is
\begin{align}
    \langle\beta_A\rangle = g_0 I_\phi(\eta) - \arctan\Big[\cos(\thg)\tan\big(g_4 I_\phi(\eta)\big)\Big]. \label{eq-D26}
\end{align}

In the extreme case that $\td\rightarrow (n+\frac{1}{2})\pi$, it is possible for $|b_1|$ to be smaller than $|c_2|$, so the transition from the zero to the nonzero branch at $a_1'=\sqrt{c_2^2-b_1^2}$ contributes an extra term to the integral 
\begin{align}
    I &= \arctan\Big(\frac{a_1}{b_1}\Big) -  \arctan\Big(\tfrac{\sqrt{c_2^2-b_1^2}}{b_1}\Big)\\ 
    \Rightarrow\ \ \langle\varphi_A^\pm\rangle&= \arctan\Big[\cos(\thg)\tan\big(\td\big)\Big]-\arctan\sqrt{\tR_0^2\sin(\thg)^2\tan\big(\td\big)^2-1}\\
    &=\mp \arctan\Big[\cos(\thg)\tan\big(g_4 I_\phi\big)\Big]- \arctan\sqrt{\tR_0^2\sin(\thg)^2\tan\big(g_4 I_\phi\big)^2-1}
\end{align}
And the birefringence angle is exactly the same as what in eq.~\eqref{eq-D26}.

\bibliographystyle{unsrtnat}
\bibliography{sample}


\end{document}